\documentclass[jgrga]{agutex}
\usepackage{amsmath}
\usepackage{graphicx}

\authorrunninghead{PRANTS ET AL.}
\titlerunninghead{IDENTIFYING LAGRANGIAN FRONTS}
\authoraddr{S. V. Prants, prants@poi.dvo.ru}

\begin{document}

\title{Identifying Lagrangian fronts with favourable fishery conditions}
\authors{S. V. Prants,\altaffilmark{1}
M. V. Budyansky,\altaffilmark{1} and M. Yu. Uleysky\altaffilmark{1}}

\altaffiltext{1}{Laboratory of Nonlinear Dynamical Systems,
Pacific Oceanological Institute of the Russian Academy of Sciences,
43 Baltiiskaya st., 690041 Vladivostok, Russia, URL: dynalab.poi.dvo.ru}

\begin{abstract}
Lagrangian fronts (LF) in the ocean delineate boundaries between surface waters
with different Lagrangian properties. They can be accurately detected in
a given velocity field by computing synoptic maps of the drift of synthetic
tracers and other Lagrangian indicators.
Using Russian ship's catch and location data
for a number of commercial fishery seasons in the region of the northwest Pacific
with one of the richest fishery in the world, it is shown statistically that the saury
fishing grounds with maximal catches are not randomly distributed over the region but located mainly along those
LFs where productive cold waters of the Oyashio Current,
warmer waters of the southern branch of the Soya Current, and waters of
warm-core Kuroshio rings converge. Computation of those fronts with the altimetric
geostrophic velocity fields both in the years with the First and Second
Oyashio Intrusions shows that in spite of different
oceanographic conditions the LF locations may serve good
indicators of potential fishing grounds. Possible reasons for saury aggregation
near LFs are discussed. We propose a mechanism of effective export of nutrient rich waters
based on stretching of material lines in the vicinity of hyperbolic objects in the ocean.
The developed method, based on identifying LFs in any velocity fields, is quite general
and may be applied to forecast potential fishing grounds for the other pelagic
fishes in different seas and the oceans.

\end{abstract}

\begin{article}
\section{Introduction}

Regions, where horizontal gradients of hydrological properties go through a maximum, are ubiquitous
in the ocean. The importance of oceanic fronts to ecosystems may be explained by the fact that
they are associated with a convergent flow with an intensified flux of nutrients. If the front is sufficiently
long-lived, populations of phyto- and zooplankton will increase attracting the other higher level organisms in the
trophical chain which are able to detect the front.

By the common opinion \citep{Owen,Olson,Bakun},
good fishing areas are often found at the boundaries of warm and cold
currents and around warm-core eddies where the energy of the physical system
is transferred in some way to biological processes. This strong physical-biological
interaction provides favourable conditions for marine organisms.
Surface convergent fronts of considerable physical and biological activity
occur in zones where different water types impinge. The reasons leading to aggregation of tuna, saury and some other
pelagic fishes at oceanic fronts are unknown in detail, but there are in the literature some speculations about
physical mechanisms providing a transfer of the energy of the physical system to biological processes
\citep{Owen,Olson,Bakun}. They include transport of nutrients, phytoplankton blooms and aggregation of other
marine organisms at fronts and eddy edges. Thus, oceanic fronts work as aggregating mechanisms
for zooplankton, the main food for pelagic fishes. SST gradients have long been the main indicators
used to find places in the ocean with rich marine resources.

In the northwest Pacific, the cold Oyashio Current flows out of the Arctic
along the Kamchatka Peninsula and the Kuril Islands and converges
with the warmer Kuroshio Current off the eastern shore of Japan.
This frontal zone is known to be one of the richest fishery in the world due to
the large nutrient content in the Oyashio waters and high tides in some areas
therein. Some people attempted to identify the favourable oceanographic
conditions for catching pelagic fishes.
The Kuroshio-Oyashio fronts have long been recognised by Japanese fishermen
to attract squids, fishes and mammals (for the earlier studies see Ref.
\citep{Uda38}). It was found that fishing grounds depend
on the location of the Oyashio fronts and vary from year to year
\citep{Fukushima79,Saitoh86,Sugimoto92,Yasuda96}. The
fishing grounds are located near shore if there exists the First
Oyashio Intrusion along the eastern coast of Hokkaido, Japan. In the other
years, the fishing grounds are located offshore where the Second Oyashio Intrusion
is formed due to the presence there of a large warm-core Kuroshio ring.
It has been shown that locations of the fishing grounds depend not only on
instantaneous and local oceanographic conditions nearby the fishing grounds
but on the conditions over the whole region.

We study here the connection of Lagrangian fronts (LFs), delineating
boundaries between surface waters with different Lagrangian properties,
and fishing grounds. To be concrete, we focus
on fishing grounds of Pacific saury ({\it Cololabis saira}), one of
the most commercial pelagic fishes in the region. The Pacific saury is a migratory pelagic fish moving in schools.
They used to pass for 5 -- 6 months distances of the order of 2500 miles from spawning grounds to feeding grounds.
They spend much of the time near the surface in nights and in deeper waters in daily time.
Pacific saury, in general,
migrate seasonally from the south to the north. In winter and spring,
spawning grounds are formed in the south, off the eastern coast of Honshu,
Japan. In spring and summer, juvenile and young saury migrate northward to the Oyashio
area. After feeding in those productive waters, adult saury migrate to the
south in the late summer. Commercial fishery begins in August and ends in December.

Based on the AVISO altimetric geostrophic velocity fields, we compute synoptic maps of zonal, meridional,
and absolute drift of synthetic tracers and the finite-time Lyapunov exponents (FTLE). Those maps with the catch data
from the Russian saury fishery overlaid allow to identify the LFs in the region
with favourable fishing conditions in the years both with the First and  Second Oyashio Intrusions.
In order to determine whether saury is actively associating with LFs or not, we
compute the frequency distributions of the distances between locations
of fishing boats with saury catches and the LFs. The comparison of those distributions with random distributions
of fishing boats over the same region for all the fishery seasons with available data
provides evidence that saury fishing locations
are not randomly distributed over the region but congregate near strong LFs.
Thus, the LFs may serve as a new indicator for potential fishing grounds.

The paper is organized as follows. Section~2 gives an introduction to
the Lagrangian approach to study transport and mixing in the ocean based on
application and elaboration of some methods from dynamical systems theory.
Section~3 details the data and methods used in the paper. We introduce there 
the notion of a LF and discuss briefly their specifics and difference from common 
oceanic fronts and Lagrangian coherent structures. Section~4 contains the illustrative examples
on identification of LFs favourable for saury fishing in the seasons with the First and Second Oyashio Intrusions in
the region to the east off Hokkaido and southern Kuril Islands coasts.
The representative drift and FTLE synoptic maps for two different oceanographic
situations in the region are demonstrated in that section. The quantitative results are presented
in section~5 where it is shown statistically that the saury
fishing grounds with maximal catches are not randomly distributed over the region but located mainly near strong LFs.
In Appendix, we describe our method to compute accurately the FTLE.

\section{Lagrangian approach to study transport and mixing in the ocean}

Motion of a fluid particle in a two-dimensional flow is the trajectory
of a dynamical system with  given initial conditions governed by the velocity
field. The corresponding advection equations are written as follows:
\begin{equation}
\frac{d x}{d t}= u(x,y,t),\quad \frac{d y}{d t}= v(x,y,t),
\label{adveq}
\end{equation}
where the longitude, $x$, and the latitude, $y$, of a passive particle
are in geographical minutes, $u$ and $v$ are angular zonal
and meridional components of the velocity expressed in minutes per day.
Even if the Eulerian velocity
field is fully deterministic, the particle's trajectories may be very
complicated and practically unpredictable. It means that a distance
between initially close fluid particles grows exponentially in time
\begin{equation}
\| \delta {\mathbf x}(t) \| = \| \delta {\mathbf x}(0) \|\, e^{\Lambda t},
\label{delta}
\end{equation}
where $\Lambda$ is a positive number, known as the maximal Lyapunov exponent,
which characterizes asymptotically the average rate
of the particle dispersion, and $\|\mathbf{x}\|$ is a norm of the vector
$\mathbf{x}=(x,y)$. It immediately follows from (\ref{delta}) that we are unable
to forecast the fate of the particles beyond the so-called predictability
horizon \citep{Koshel06}
\begin{equation}
T_p\simeq\frac{1}{\Lambda}\ln\frac{\|\Delta \|}{\|\Delta (0)\|},
\label{horizon}
\end{equation}
where $\|\Delta  \|$ is the confidence interval of the particle location
and $\|\Delta (0)\|$ is a practically inevitable inaccuracy in
specifying the initial location. The deterministic dynamical system
(\ref{adveq}) with a positive maximal Lyapunov exponent for almost all
vectors $\delta \mathbf{x} (0)$ (in the sense of nonzero measure) is called
chaotic. It should be stressed that the dependence of the predictability
horizon $T_p$ on the lack of our knowledge of exact location is logarithmic,
i.~e., it is much weaker than on the measure of dynamical instability
quantified by $\Lambda$. Simply speaking, with any reasonable degree of
accuracy on specifying initial conditions there is a time interval beyond
which the forecast is impossible, and that time may be rather short for
chaotic systems.

Since the phase plane of the two-dimensional dynamical system
(\ref{adveq}) is the physical space for fluid particles, many abstract
mathematical objects from dynamical systems theory (fixed points,
Kolmogorov--Arnold--Moser tori, stable and unstable manifolds, periodic and chaotic orbits, etc.)
are material surfaces, curves and points in fluid flows (for a review see
\citep{Wiggins05,Koshel06}). It is well known that
besides ``trivial'' elliptic fixed points, the motion around which is stable, there
are hyperbolic fixed points which organize fluid motion in their neighbourhood
in a specific way. In a steady flow the hyperbolic points are typically
connected by the separatrices which are their stable and unstable
invariant manifolds. In a time-periodic flow the hyperbolic points are replaced
by the corresponding hyperbolic trajectories with associated invariant manifolds
which in general intersect transversally resulting in a complex manifold
structure known as homo- or heteroclinic tangles. The fluid motion in these regions
is so complicated that it may be strictly called chaotic,
the phenomenon known as chaotic advection (for a review see \citep{Koshel06}).
Adjacent  fluid particles in such tangles rapidly diverge providing
very effective mechanism for mixing.

Stable and unstable manifolds
are important organizing structures in the flow because they attract and repel
fluid particles (not belonging to them) and
partition the flow into regions with quantitatively different types of motion.
Invariant manifold in a two-dimensional flow is a material line, i.~e.,
it is composed of the same fluid particles in course of time.
By definition \citep{Wiggins05,Koshel06}, stable  ($W_s$) and unstable ($W_u$) manifolds of a hyperbolic
trajectory $\gamma(t)$~are material lines consisting of a set of points
through which at time moment $t$ pass trajectories that are asymptotical to
$\gamma(t)$ at $t \to \infty$ ($W_s$) and $t \to -\infty$ ($W_u$).
They are complicated  curves infinite in time and space that act
as boundaries to fluid transport.

The real oceanic flows are not, of course, strictly time-periodic.
However, in aperiodic flows there exist under some
mild conditions hyperbolic points and trajectories of a transient nature.
In aperiodic flows it is possible to identify aperiodically
moving hyperbolic points with stable and unstable effective manifolds
\citep{Haller00}. Unlike the manifolds in steady and periodic flows, defined
in the infinite time limit, the ``effective'' manifolds of aperiodic
hyperbolic trajectories have a finite lifetime. The point is that they
play the same role in organizing oceanic flows as do invariant manifolds
in simpler flows \citep{Haller,Haller00,H00}. The effective manifolds in course of their life undergo
stretching and folding at progressively small scales and intersect each other
in the homoclinic points in the vicinity of which fluid particles move
chaotically. Trajectories of initially close fluid particles diverge rapidly
in these regions, and particles from other regions appear there. It is the
mechanism for effective transport and mixing of water masses in the ocean.
Moreover, stable and unstable effective manifolds constitute Lagrangian
transport barriers between different regions because they are material
invariant curves that cannot be crossed by purely advective processes.

The stable and unstable manifolds of influential hyperbolic trajectories
are so important because (1) they form a kind of a template around which
oceanic flows are organized, (2) they divide a flow in dynamically different regions,
(3) they are in charge of forming an inhomogeneous mixing with
spirals, filaments and lobes, (4) they are transport barriers
separating water masses with different characteristics.
Stable manifolds act as repellers for surrounding waters
but unstable ones are attractors. That is why unstable manifolds may be
rich in nutrients being oceanic ``dining rooms''.

To quantify chaos in dynamical systems, one computes the maximal
Lyapunov exponent which enables to detect and
visualize stable and unstable manifolds as well. In aperiodic flows, it is instructive
to use with these aims the FTLE that is the finite-time average of the maximal separation rate for a pair of
neighbouring advected particles. This quantity can be computed accurately by the method introduced in Ref.~\citep{OM11}.

\section{Data and Methods}

The general method we used
is based on the Lagrangian approach to study mixing and transport at the sea
surface \citep{Haller,Haller00,H00,Boffetta01,Mancho04,Ovidio04,Shadden05,
Kirwan06,Lehahn07,Beron08,OM11,DAN11,FAO13} when one follows fluid particle trajectories in a velocity
field calculated from altimetric measurements or obtained as an output of
one of the ocean circulation model. The important notion in that approach is
so-called Lagrangian coherent structures (LCS) \citep{Haller,Haller00,H00}.
The well-known coherent structures  in the ocean are eddies and jet currents that can be visible in Eulerian velocity
fields and at satellite images of the SST and/or chlorophyll concentration. The LCSs, in general, are not visible
at snapshots but can be computed with a given velocity field by special methods.
Lagrangian coherent structures are operationally
defined as local extrema of the scalar FTLE field \citep{Haller,Haller00,H00}.
They are the most influential attracting and repelling
hyperbolic material surfaces which are curves in 2D velocity fields. The LCS are Lagrangian because they are invariant
material curves consisting of the same fluid particles. They are coherent because they are comparatively long lived
and more robust than the other adjacent structures.  The LCS are connected with the
effective stable and unstable invariant manifolds of hyperbolic (unstable) trajectories that can be approximately
identified with the help of the FTLE and other Lagrangian indicators, hyperbolicity time \citep{Haller00},
patchiness \citep{Lukovich}, the so-called M-function which measures
the Euclidean arc-length of the curves outlined by trajectories for a finite-time interval \citep{Mancho2009},
the correlation dimension of trajectory and the so-called ergodicity defect \citep{Rypina11}.
A new approach to locating material transport barriers in unsteady planar fluid flows has been introduced
recently \citep{H11,H12}. Seeking transport barriers as minimally stretching material lines,
it is possible to locate hyperbolic barriers (generalized stable and unstable manifolds), elliptic barriers
(generalized KAM curves) and parabolic barriers (generalized shear jets) in temporally aperiodic flows
defined over a finite time interval. This approach \citep{H11,H12} also yields a metric (geodesic deviation)
that determines the minimal computational time scale needed for a robust numerical identification of generalized LCSs.

The convenient measures to distinguish trajectories with different particle's
fate and origin have been introduced recently, namely, the time particles need to leave a given region \citep{OM11},
absolute, zonal, and meridional displacements of particles from their initial positions \citep{DAN11,FAO13}.
The impact of LCSs on biological organisms has been recently studied in
Ref.~\citep{Kai09}. By comparing the seabird satellite positions with computed LCSs locations,
it was found that a top marine predator, the Great Frigatebird, was able to track the LCSs in the Mozambique Channel
identified with the help of the finite-size Lyapunov exponent.

The LCSs  are determined mainly by the large-scale advection field, which is appropriately captured by
altimetry. It has been shown theoretically \citep{H02} that the LCSs are robust to errors in
observational or model velocity fields if they are strongly attracting or repelling and exist
for a sufficient long time. This is due to the fact that though the particle
trajectories will in general diverge exponentially from the true trajectories near a
repelling LCS, the very LCSs are not expected to be perturbed to the same degree, because errors in the particle
trajectories spread along the LCS. A few numerical experiments have been carried out to test the sensitivity of LCSs,
found by identifying ridges of the FTLE, to errors in altimetric velocity fields.
LCS were found to be relatively insensitive to both sparse spatial and temporal resolution and to
the velocity field interpolation method \citep{Harrison10,Keating11,Hernandez11}. The FTLE
method is reliable for locating boundaries of large eddies and strong jets. However, small LCS features
are not well resolved from altimetry and should be considered with some caution.
The comparison of the LCSs computed with altimetric velocity fields in numerous paper
(see, for example, \citep{Abraham02,Olascoaga06,
Beron08,Ovidio09,Huhn12}) with independent satellite and in situ measurements of thermal and other fronts in different
seas and oceans has been shown a good correspondence.

Our approach is based on searching for specific Lagrangian features in the altimetric geostrophic velocity fields which 
indicate to the presence of convergence of waters with different properties.
We call them Lagrangian Fronts (LF), which are boundaries between surface waters with different Lagrangian properties. 
It may be, for example, a thermodynamical property, such as SST, salinity, density, etc. or concentration of chlorophyll-a. 
Lateral maximal gradients of those properties would indicate on common oceanic fronts, thermal, salinity, density and 
chlorophyll ones, which are often connected with each other. However, one may consider more specific Lagrangian 
indicators which are functions of a particle's trajectory,  
such as a distance passed for a given time, absolute, meridional, and zonal displacements of particles
from their initial positions, the numbers of their cyclonic and anticyclonic rotations, etc. Even in the situation where 
the water itself is indistinguishable, say, in temperature, and the corresponding SST image does not show a thermal front 
there may exist a LF separating waters with the other distinct properties. 

Common oceanic fronts are manifestations of the current state of a water medium that can be detected by direct 
measurements. SST and ocean color fronts are visible on satellite images which, however, are not available in cloudy or 
rainy days. The LFs reflect history and origin of water masses and can be computed and visualized on Lagrangian maps. 
They can be, in principle, measured by launching a large number of drifters in appropriate places. 
The LFs may coincide with common oceanic fronts but may not. It is possible to compute the LF of such a Lagrangian 
indicator that would not manifest itself as an oceanic front but would give a useful information on water motion. 
Even if a specific LF does not coincide with a common oceanic front, it does not mean that it would be usefulness. 
Such the LF may simply reflect the other properties of convergent water masses 

The relationship between LCSs and LFs is not trivial. Any LF by definition is a curve with the maximal local gradient of 
a Lagrangian property which varies significantly on both sides of the LF, whereas the FTLE values are almost the same on 
both sides of any ridge in the FTLE field defining a LCS. By definition, any Lagrangian indicator is a function of trajectory, 
whereas in order to compute the FTLE it is necessary in addition to know the dynamical system as well. 
LCSs contain information about evolution of some medium segments. The drift and the other Lagrangian indicators are 
characteristics of a given fluid particle whereas the Lyapunov exponent is a characteristic of the medium surrounding 
that particle. Moreover, LCSs and LFs have a different topology on Lagrangian maps, the first ones are 2D bands 
(ridges) whereas the latter ones are curves (gradients). We would like to stress the important role of the LFs because, 
in difference from rather abstract geometric objects of an associated dynamical system, like invariant manifolds, $W_s$ 
and $W_u$, they are fronts of real physical quantities that can be, in principle, measured.

Of particular importance in detecting LFs is the drift, $D$, that is simply the distance between the final, $(x_f,y_f)$, and
initial, $(x_0,y_0)$,  positions of advected particles on the Earth sphere with the radius
$R$
\begin{equation}
D\equiv R\arccos[\sin y_0 \sin y_f +\cos y_0 \cos y_f \cos (x_f - x_0)],
\label{drift}
\end{equation}
This quantity and zonal, $D_x$, and meridional, $D_y$,  drifts have been shown to be useful
in quantifying transport of radionuclides in the Northern Pacific
after the accident at the Fukushima atomic plant station \citep{DAN11} and transport
of the Madagascar plankton bloom \citep{Huhn12}.

The FTLE field characterizes quantitatively mixing along with directions of maximal stretching and contracting,
and it is applied to identify LCSs in irregular velocity fields. The FTLE is computed here by the method
of the singular-value decomposition of an evolution matrix for the linearized advection equations \citep{OM11}
and is given by formulae (for details see Appendix)
\begin{equation}
\lambda(t,t_0)=\frac{\ln\sigma(t,t_0)}{t-t_0},
%\label{lyap_ftle}
\end{equation}
which is the ratio of the logarithm of the maximal possible
stretching in a given direction to a time interval $t-t_0$. Here $\sigma(t,t_0)$ is the maximal singular value of
the evolution matrix.
The method proposed enables to compute accurately the FTLE in altimetric velocity fields.

Geostrophic velocities  were obtained from the AVISO database (http://www.aviso.oceanobs.com). The data is gridded on a $1/3^{\circ}\times
1/3^{\circ}$ Mercator grid.
Bicubical spatial interpolation and third order Lagrangian polynomials in time have been used to provide
accurate numerical results. Lagrangian trajectories have been computed by
integrating the advection equations with a fourth-order Runge-Kutta scheme with a fixed time step of $0.001$th part of a day.

The SST data (http://oceancolor.gsfc.nasa.gov) were used to illustrate oceanographic conditions in the cases of the First and Second Oyashio
Intrusions. The data on fishing positions in latitude and longitude and daily
catches were obtained from the database of the Federal Agency for Fishery of the
Russian Federation.

\section{Results}

\subsection{Identification of the Lagrangian fronts favourable for saury
fishing in the season with the First Oyashio Intrusion}

We restrict our analysis in this paper by the region to the east off
Hokkaido (Japan) and southern Kuril Islands (Russia) coasts where the daily saury catch data from
the Russian fishery were collected for a few fishery seasons.
With the aim to detect the LFs separating waters with different origin and histories,
we distribute a large number of synthetic particles over that region and integrate advection equations (\ref{adveq})
backward in time for two weeks to compute the zonal, $D_x$, meridional, $D_y$, and absolute, $D$,
particle's displacements from their positions on the initial day. Coding the particle's displacements by color,
one gets the synoptic maps that provide the evidence of origin of water masses present in the region on a given day.

\begin{figure*}[htb]
\begin{center}
\includegraphics[width=0.35\textwidth]{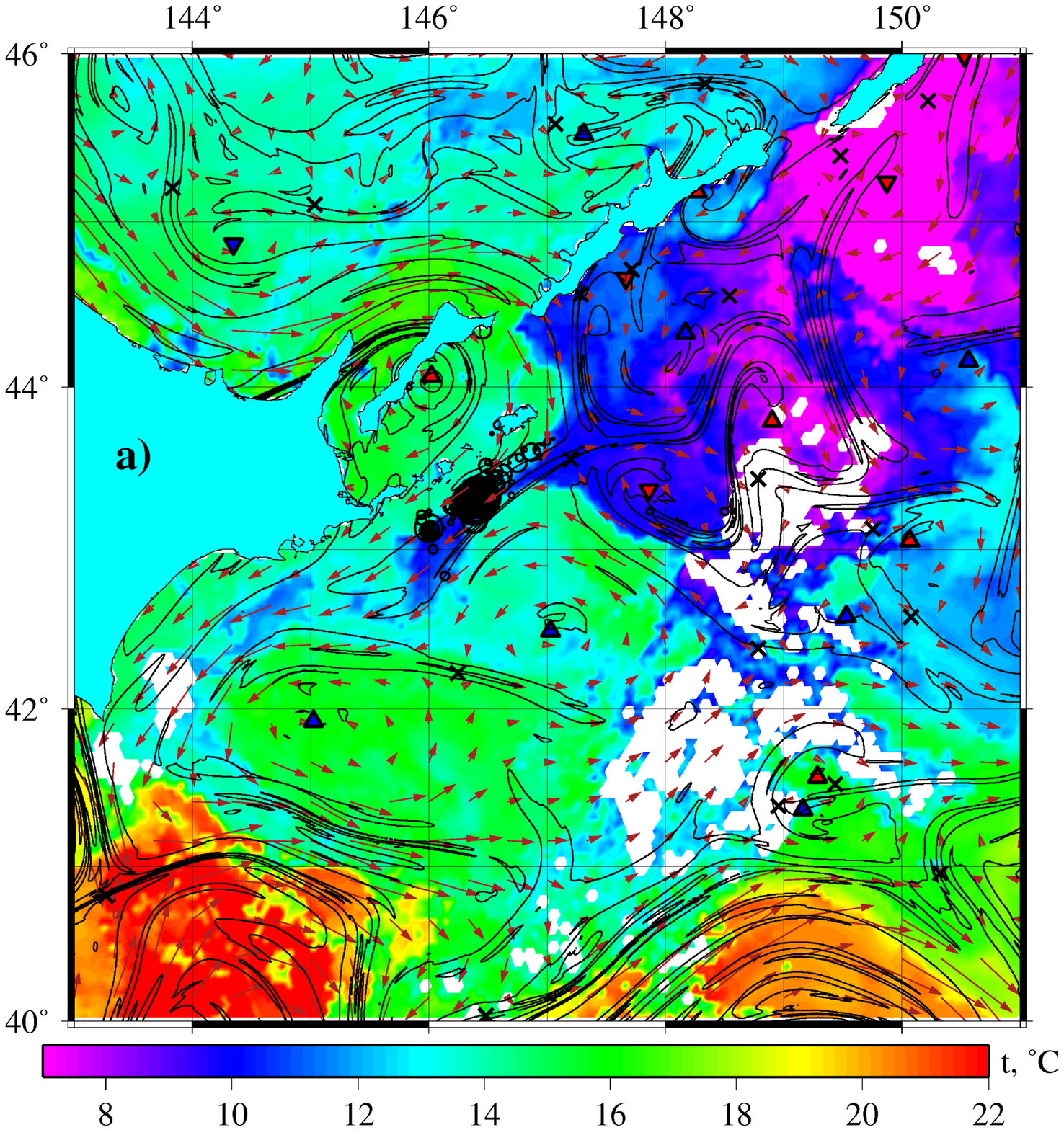}
\includegraphics[width=0.35\textwidth]{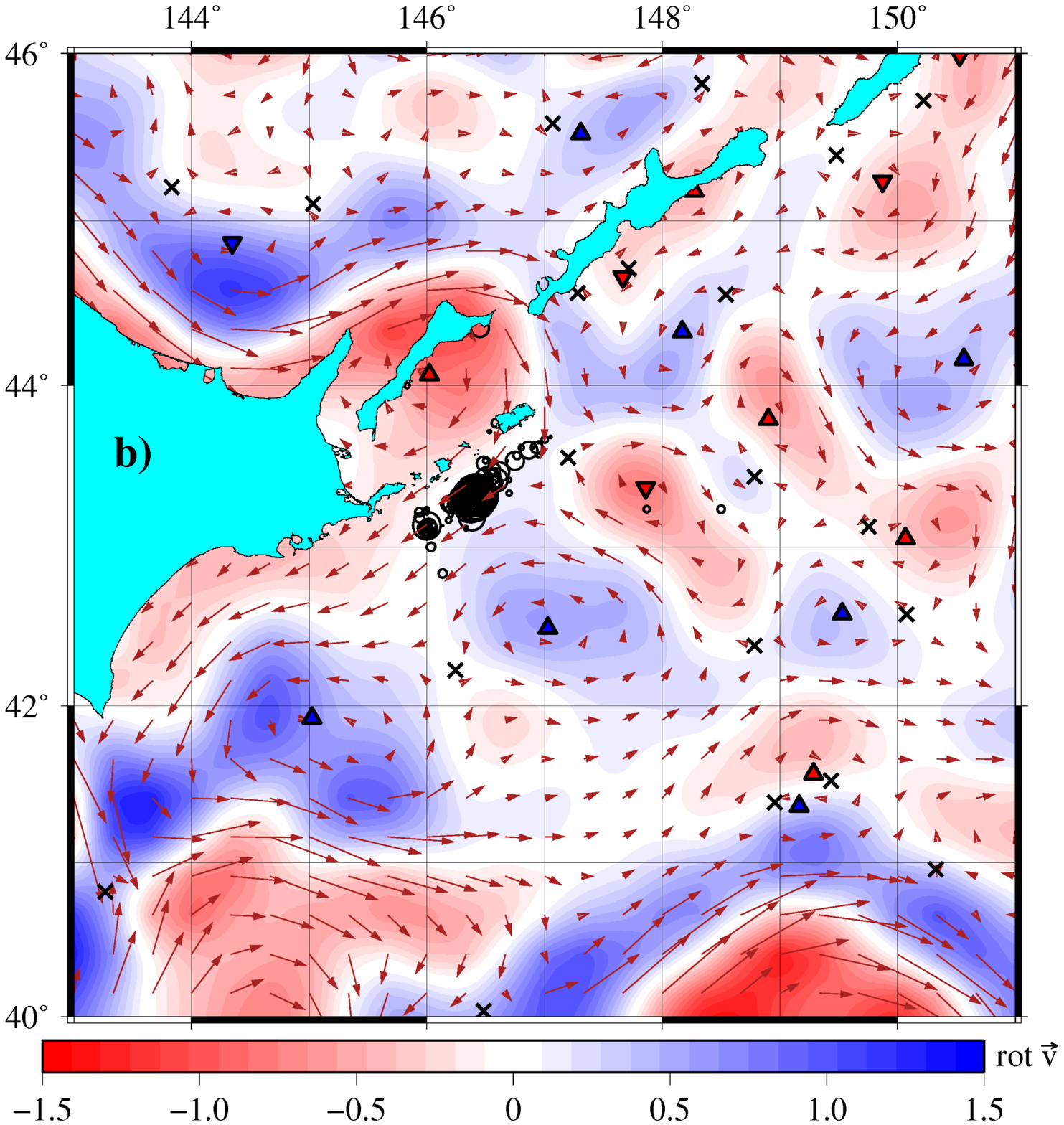}\\
\includegraphics[width=0.35\textwidth]{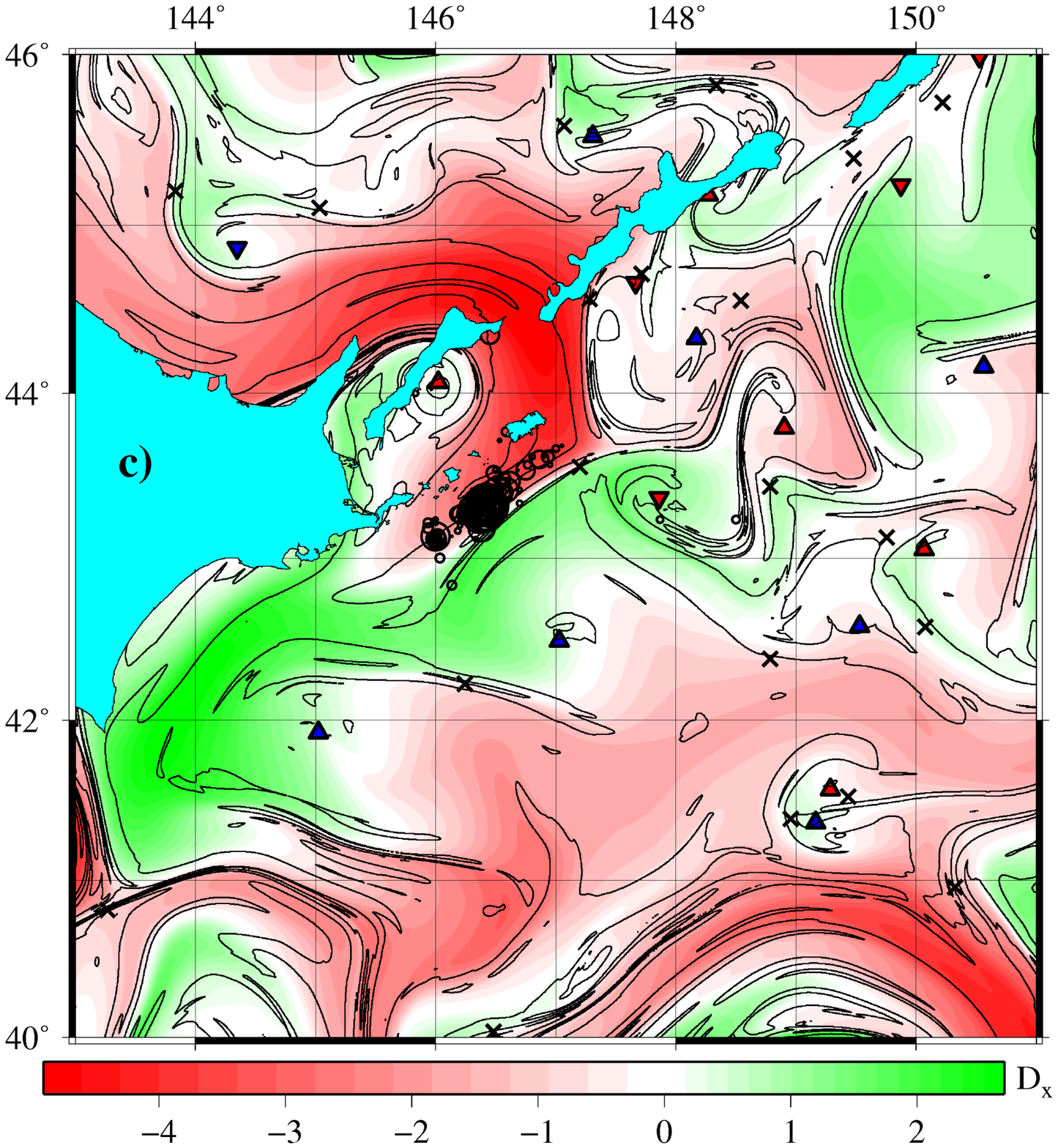}
\includegraphics[width=0.35\textwidth]{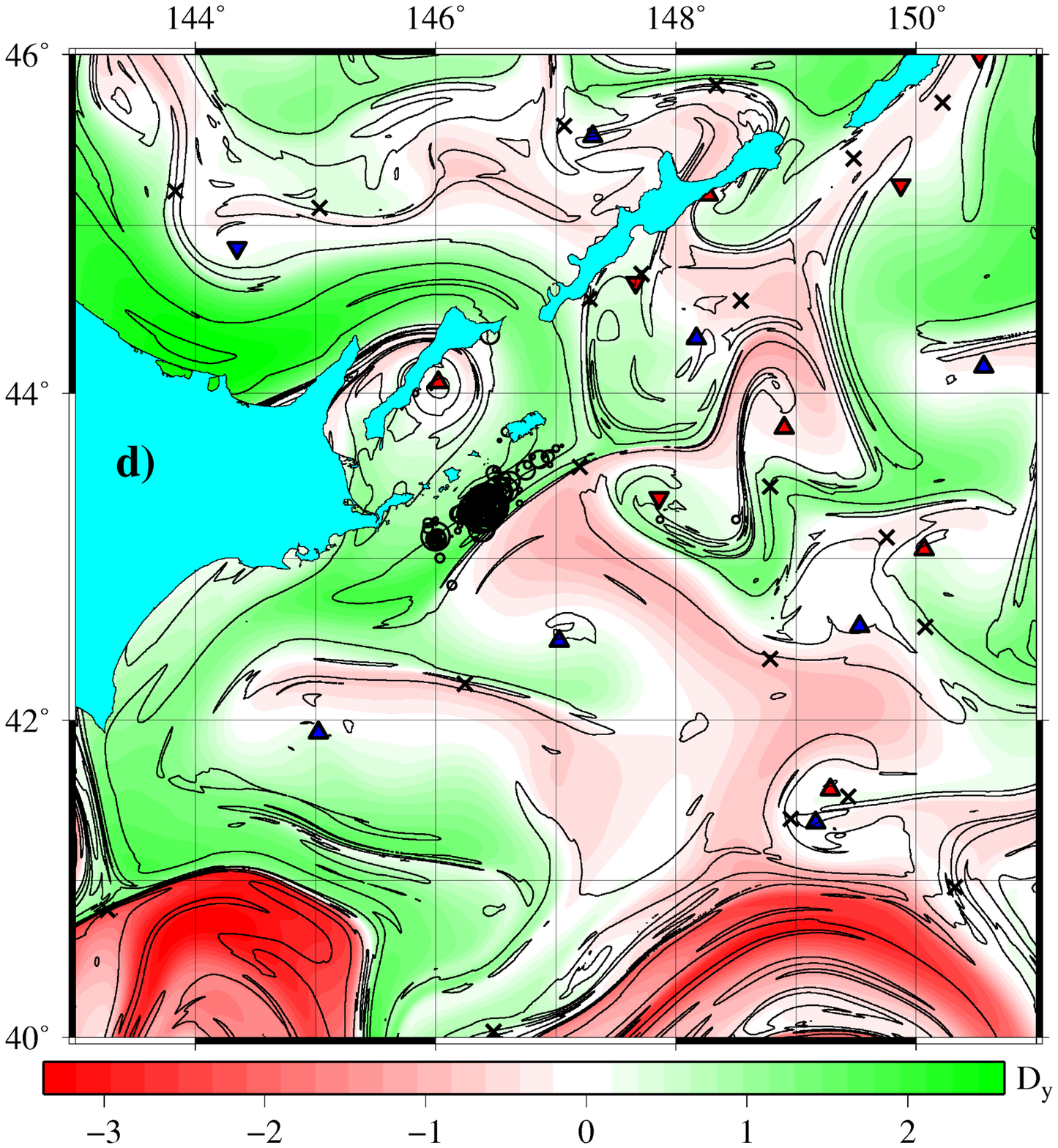}\\
\includegraphics[width=0.35\textwidth]{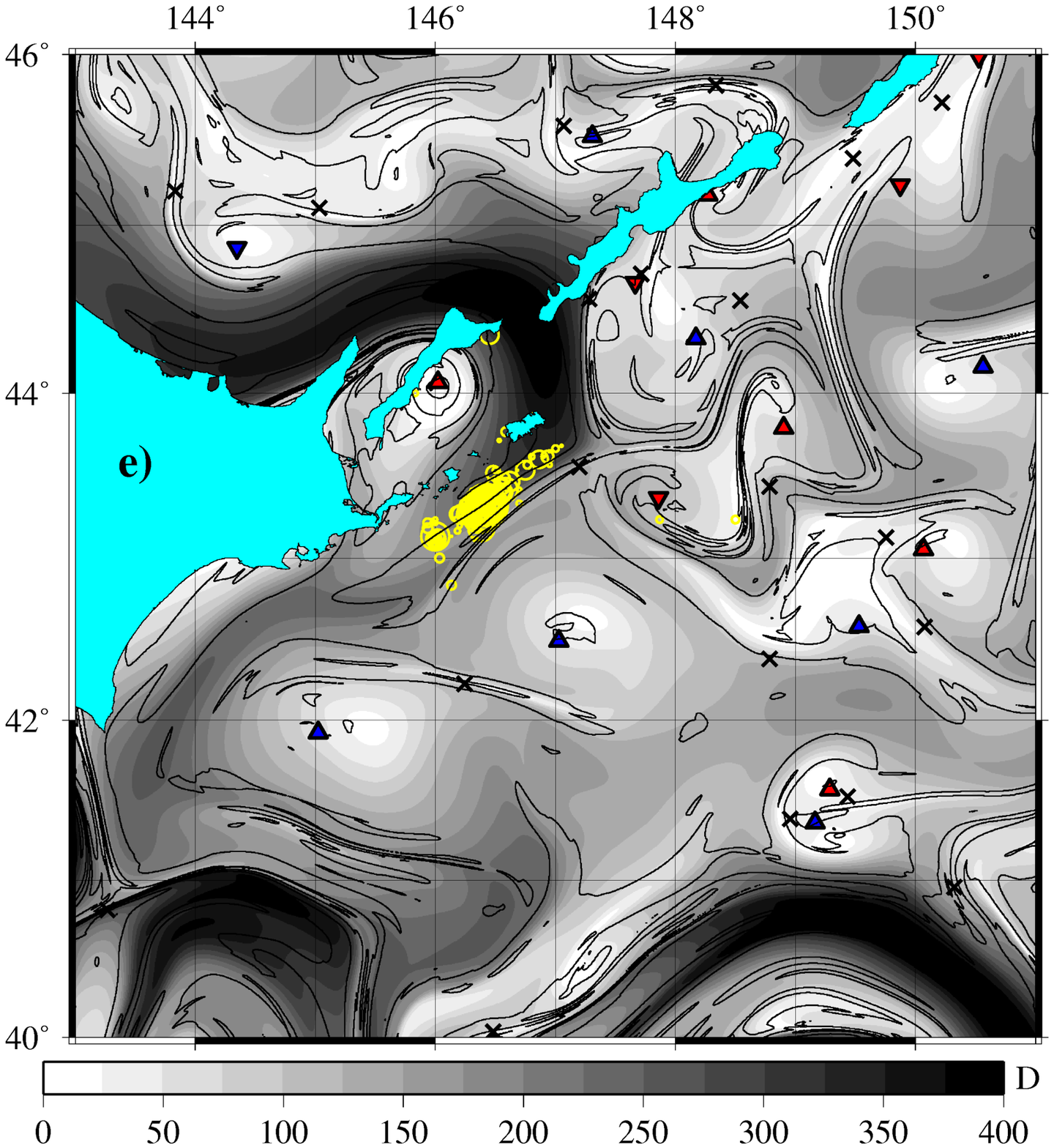}
\includegraphics[width=0.35\textwidth]{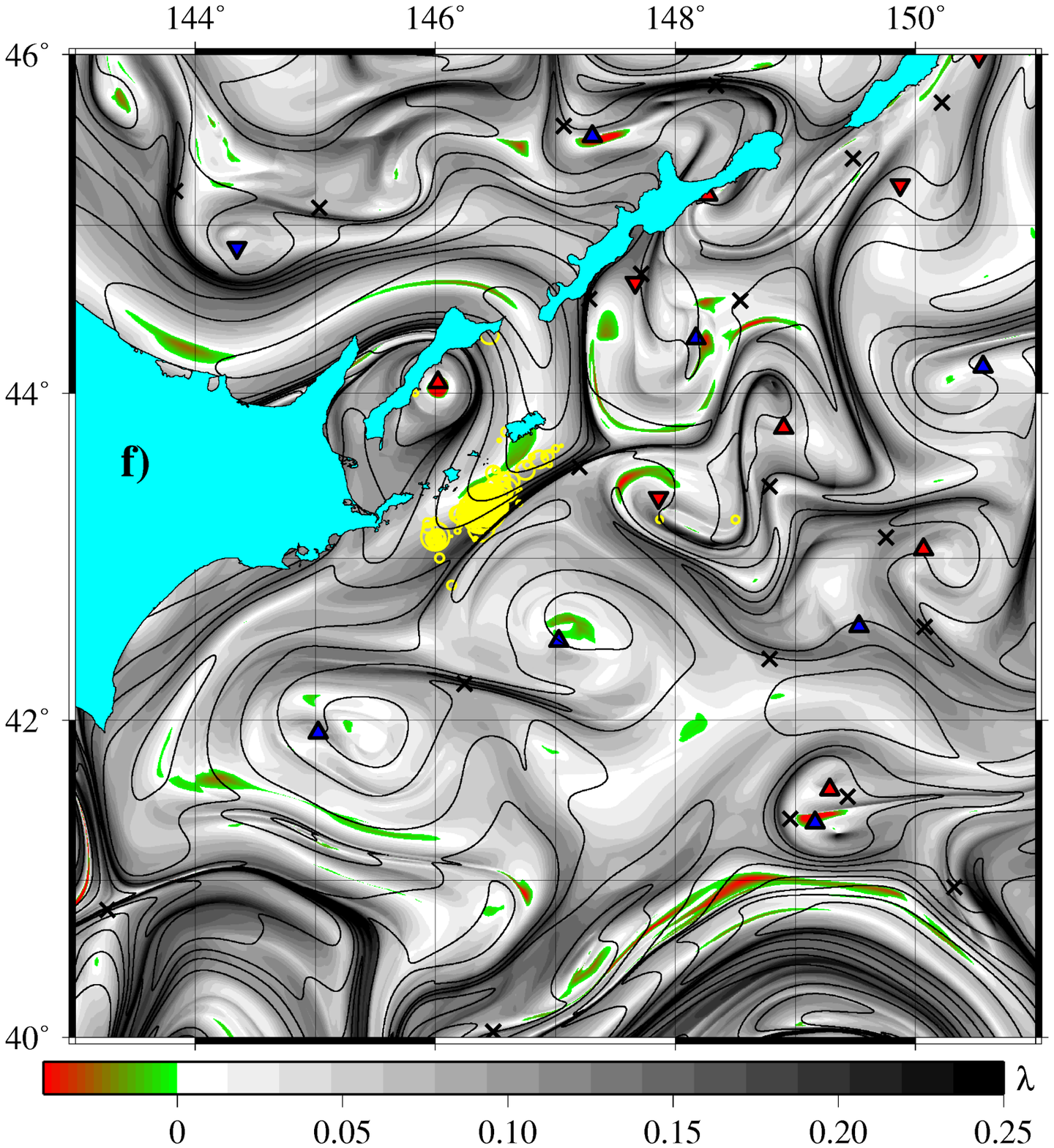}
\end{center}
\caption{The season with the First Oyashio Intrusion.
(a) SST image, (b) vorticity map, (c) zonal, (d) meridional, and (e) absolute
drift maps, and (f) FTLE map on 24 September 2002 with locations of saury catches imposed.
FTLE is in units of $[\rm days]^{-1}$, the zonal and meridional drifts, $D_{x,y}$,
are in degrees, and the absolute drift, $D$, is in km.
See explanation in the text.}
\label{fig1}
\end{figure*}

In some seasons there exists the First Oyashio Intrusion along the eastern
coast of Hokkaido which is visible on the SST image in
Fig.~\ref{fig1}a averaged for 23--25 September 2002. It is a combined map with
the velocity field specified by arrows, contour lines of the FTLE, $\lambda$,
and locations of ``instantaneous'' hyperbolic (crosses) and elliptic (triangles)
fixed points imposed. Blue and red triangles mean cyclonic and anticyclonic rotations,  respectively. 
The data of saury catch for 23--25 September 2002 overlaid allow to conclude that the
fishery grounds have been concentrated in the waters of the
First Oyashio Intrusion (the radius of the circles in the figure
is proportional to the catch in tons per a given ship).
In Fig.~\ref{fig1}b the vorticity field, $\operatorname{rot} \mathbf{v}$, together with
the velocity field, $\mathbf{v}=(u,v)$, are shown on 24 September 2002 with the locations
of ``instantaneous'' hyperbolic (crosses) and elliptic (triangles)
fixed points imposed. The region contains a number of the vorticity patches with
cyclonic (blue) and anticyclonic (red) rotations.

Computing backward in time the displacements
for a large number of particles from their initial to final positions,
we get the zonal, $D_x$, and meridional, $D_y$, drift maps with the contour lines of the
FTLE and the black circles of saury catch locations imposed. Those maps
visualize clearly the LFs with convergent water
masses with different origin and histories.
The color on the zonal drift map (Fig.~\ref{fig1}c) distinguishes
the particles entering to the region through its
western (red) and eastern (green) boundaries
whereas red and green colors on the meridional
drift map (Fig.~\ref{fig1}d) mean that particles enter
to the region through its southern and northern boundaries, respectively.
Nuances of the color on both the figures code the
distance passed by the corresponding particles in the zonal or meridional
directions in degrees.
It is seen on both the drift maps that
productive cold waters of the Oyashio Current and warmer waters of the
southern branch of the Soya Current flowing through the straights between
the southern Kuril Islands converge at the LFs with the maximal catches.
This LF demarcates the boundary between the ``green'' and ``red'' waters.
The absolute drift map, $D(x,y)$, in Fig.~\ref{fig1}e confirms this conclusion
visualizing the same LF with maximal catches (yellow circles) as a boundary between
``dark'' and ``grey'' waters.

In Fig.~\ref{fig1}f we show the FTLE field on
24 September 2002, computed by the method described in Appendix,
with the contours of the particle's absolute drift $D$.
The black ``ridges'' (curves of the local maxima) of that field are known to
delineate stable manifolds of the most influential hyperbolic trajectories
in a region when integrating advection equations forward in time
and unstable ones when integrating them backward in time.
The black ``ridges'' on the map in Fig.~\ref{fig1}f delineate the corresponding
unstable manifolds which are by definition the curves of maximal stretching.
The adjacent particles, belonging to them, may have a different history because
the larger the FTLE value the larger is an initial distance between the corresponding particles (backward-in-time integration).
The particles with maximal FTLE came to their locations from very different places.
Thus, the black ``ridges'' of the FTLE field demarcate approximately 
the corresponding LFs
(compare Figs.~\ref{fig1}e and f). On the other hand, there exist the LFs 
that are not connected with FTLE ``ridges''.
Animation of the daily Lagrangian maps for August~-- December 2002
with the fishery grounds overlaid is available at http://dynalab.poi.dvo.ru/data/GRL12/2002.
\begin{figure*}[htb]
\begin{center}
\includegraphics[width=0.33\textwidth]{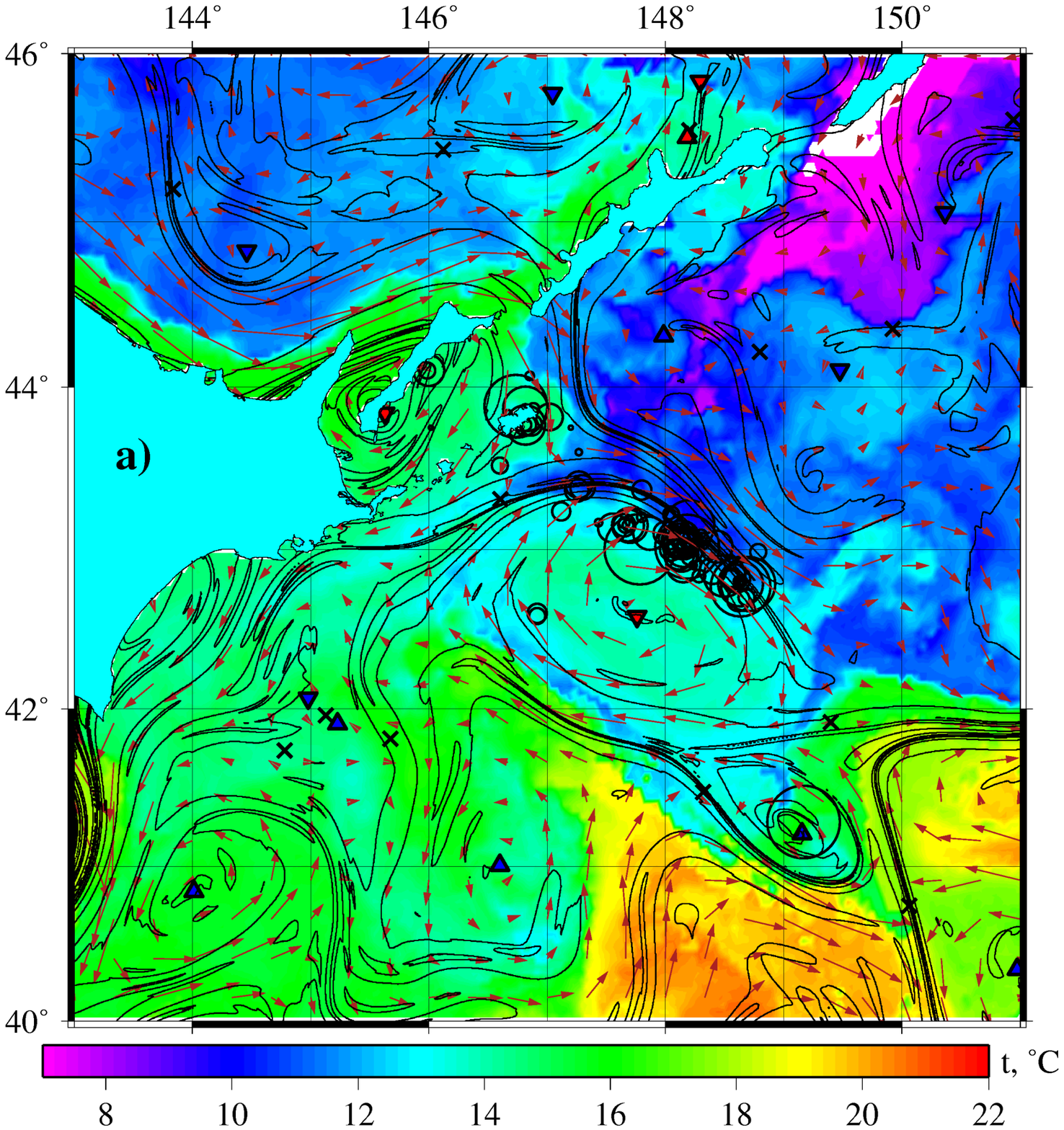}
\includegraphics[width=0.33\textwidth]{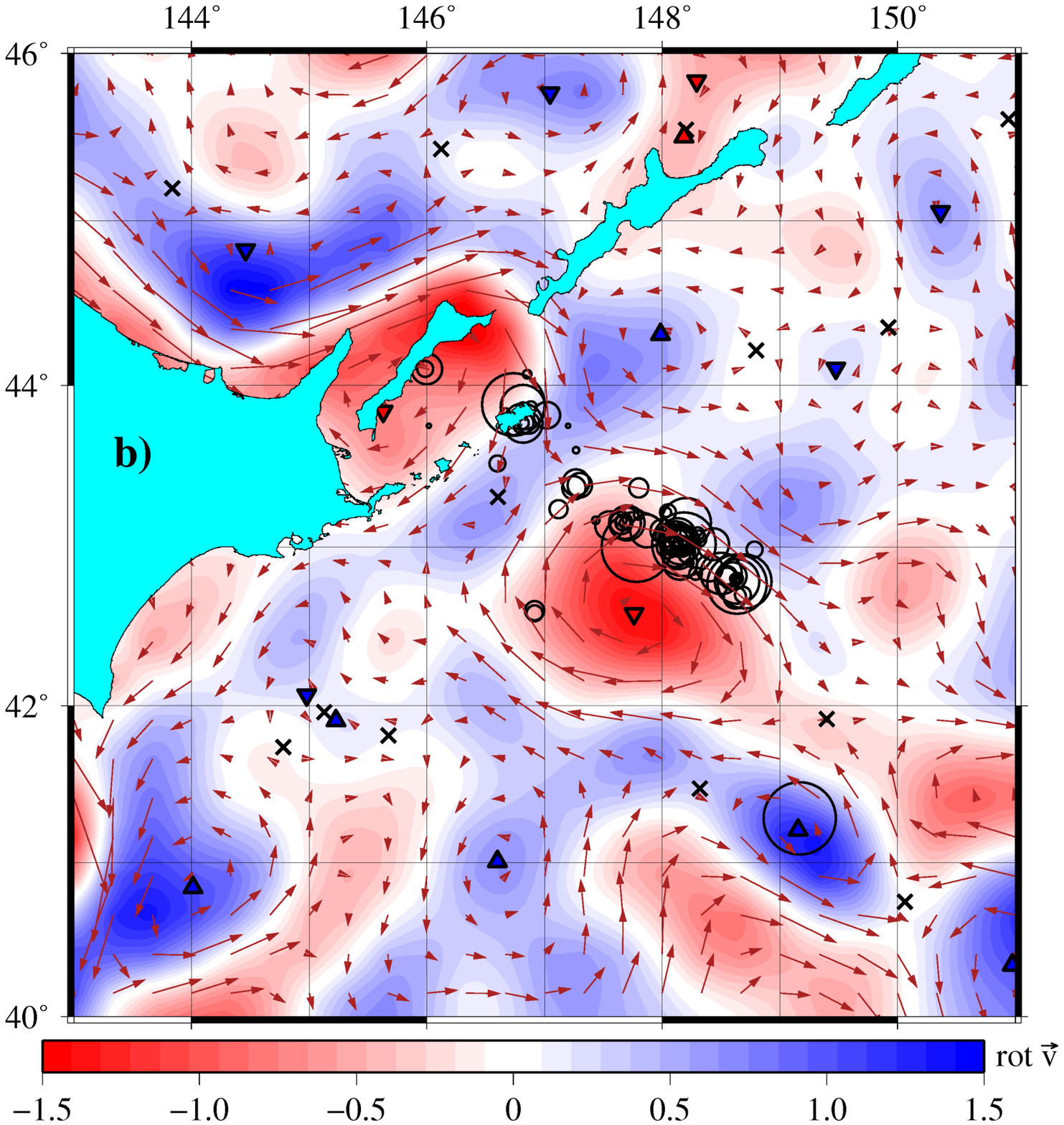}\\
\includegraphics[width=0.33\textwidth]{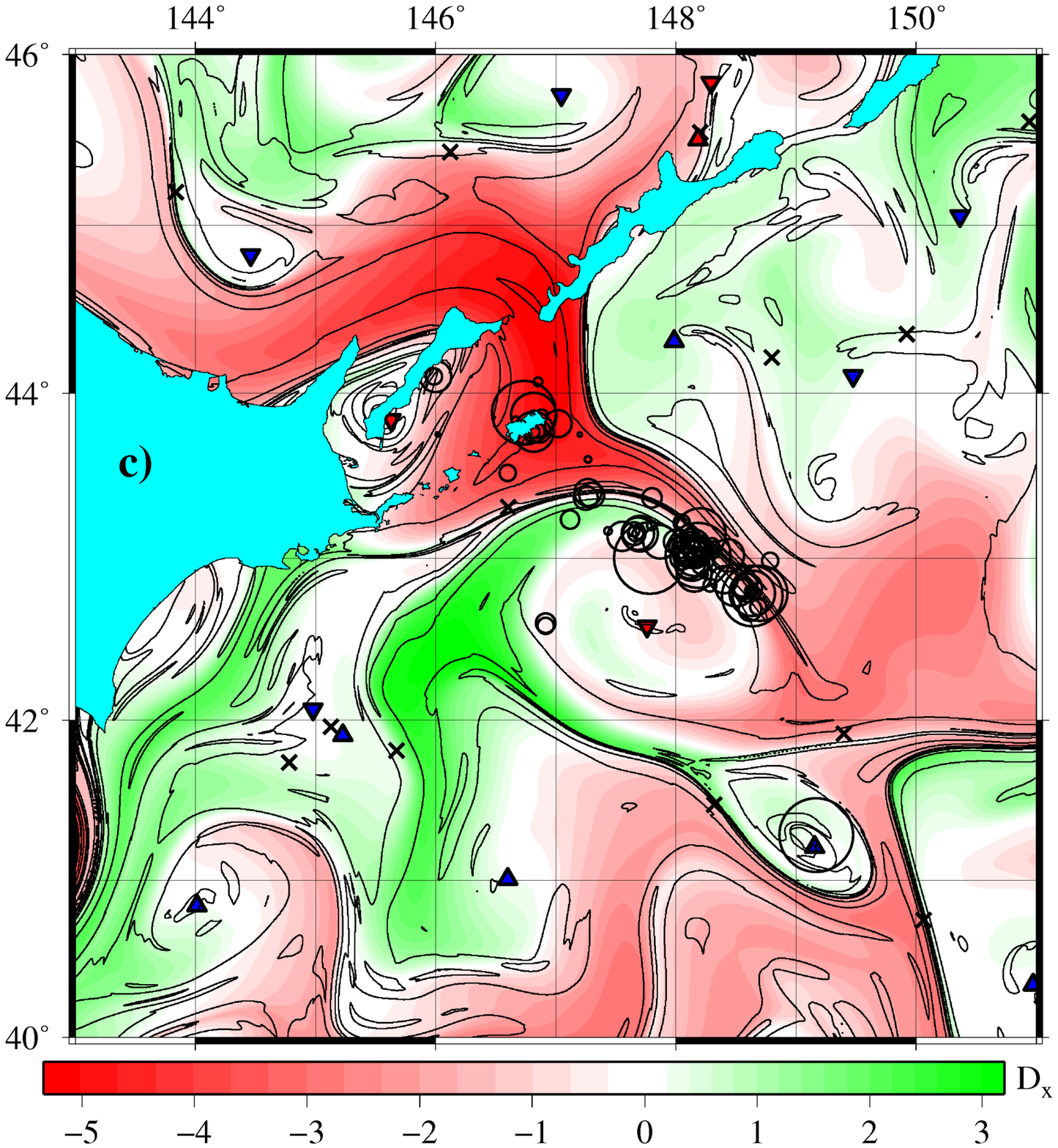}
\includegraphics[width=0.33\textwidth]{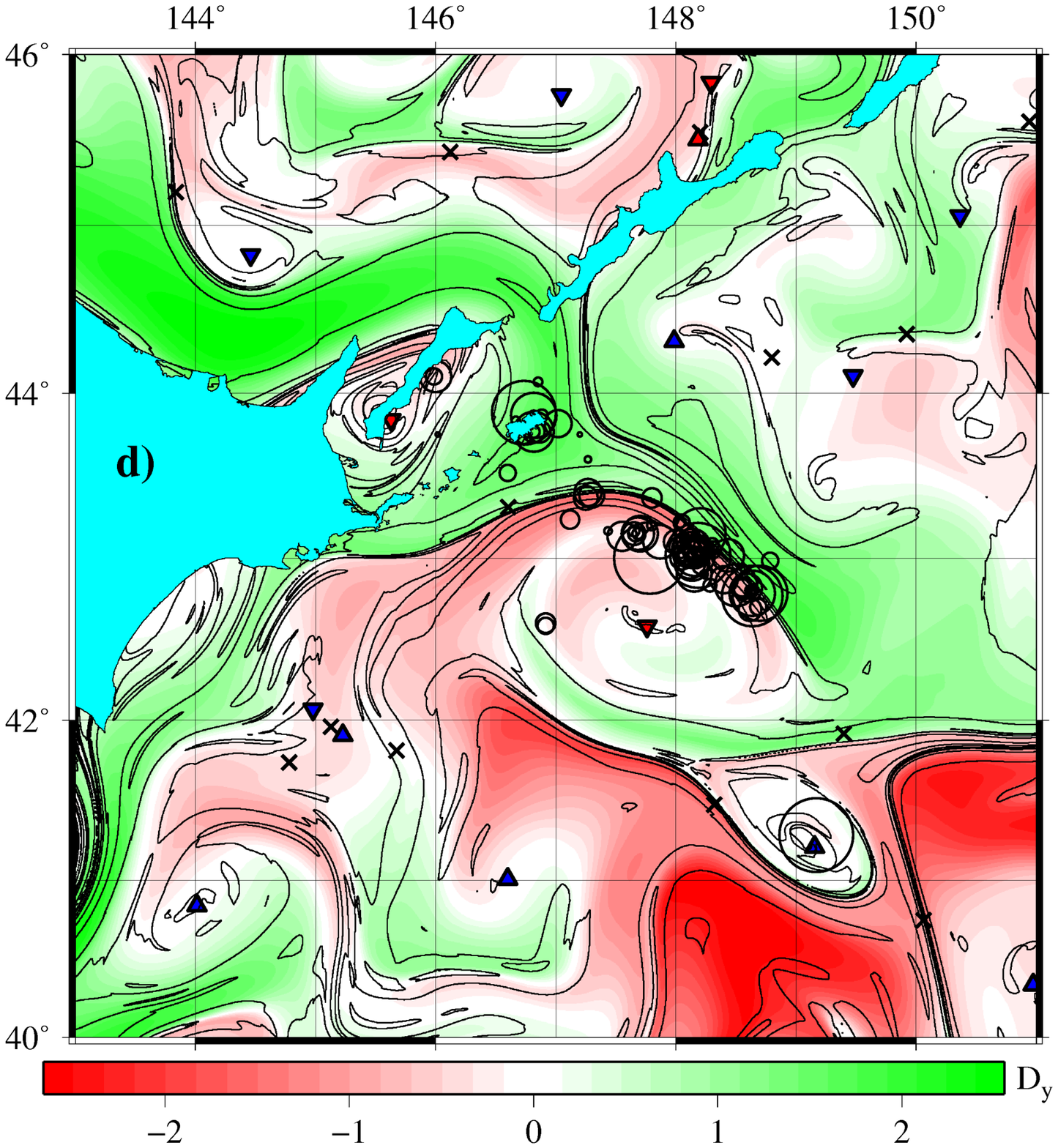}\\
\includegraphics[width=0.33\textwidth]{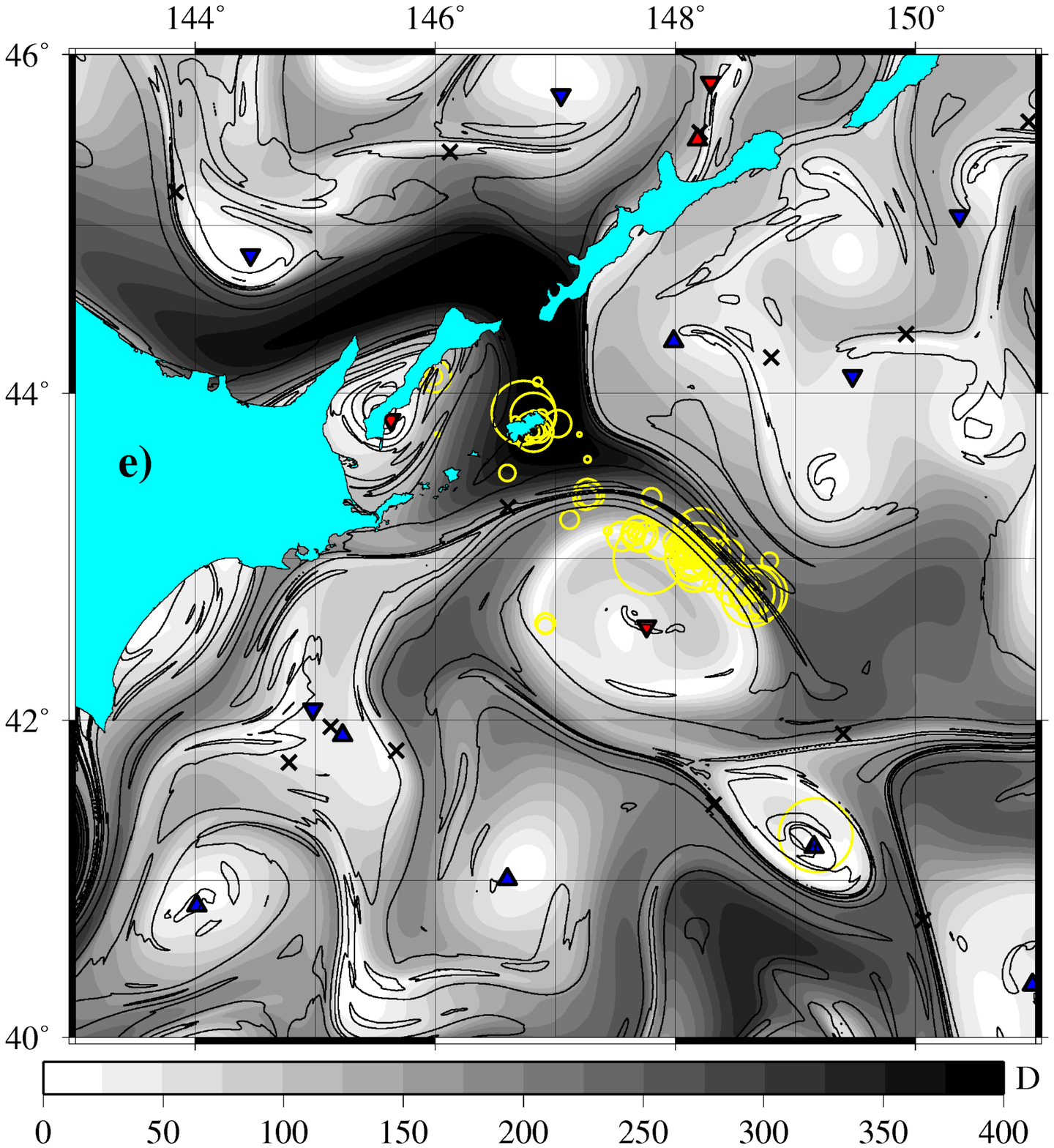}
\includegraphics[width=0.33\textwidth]{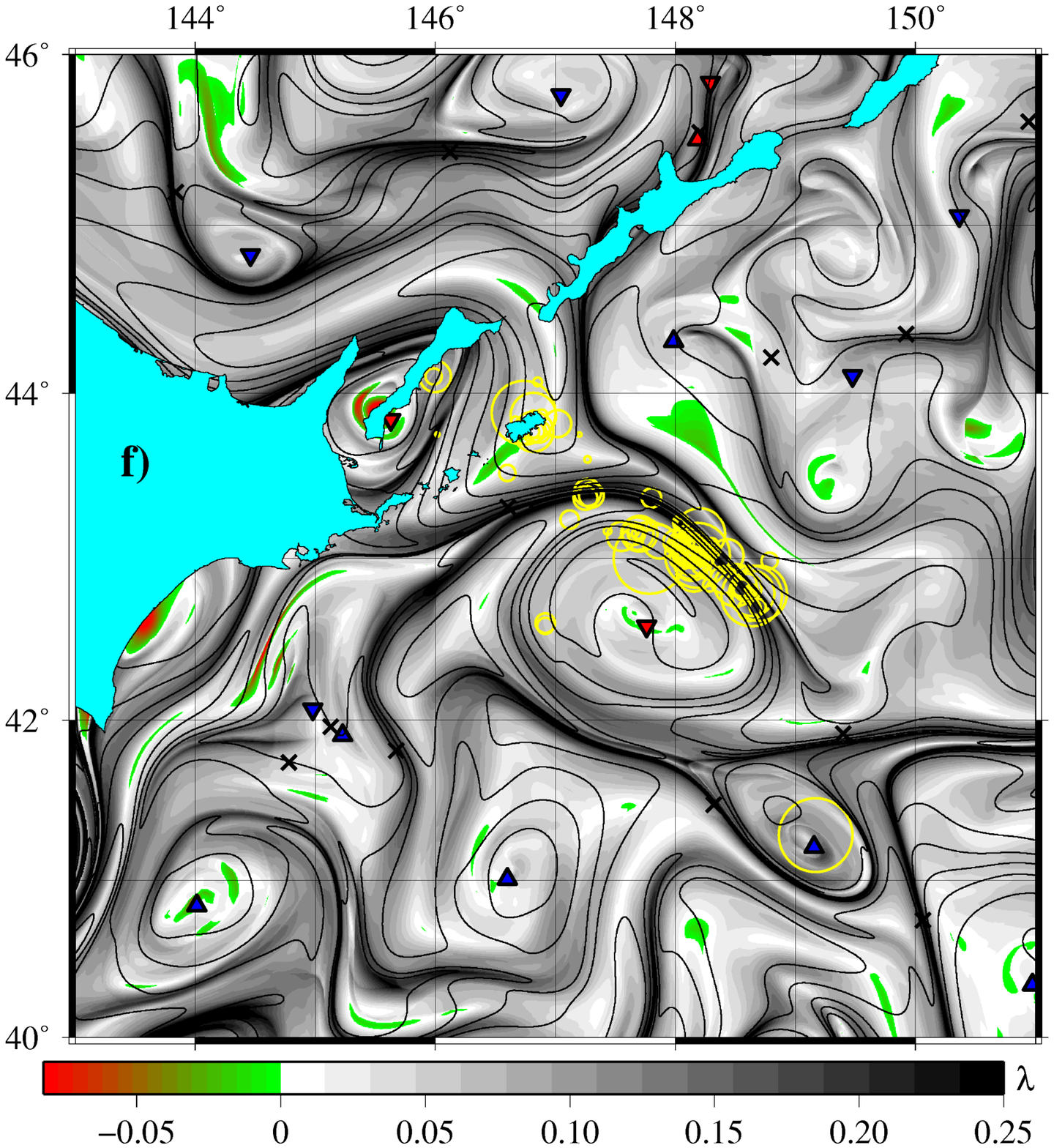}
\end{center}
\caption{The season with the Second Oyashio Intrusion.
(a) SST image, (b) vorticity map, (c) zonal, (d) meridional, and (e) absolute
drift maps, and (f) FTLE map  on 17 October 2004 with locations of maximal saury catches imposed.}
\label{fig2}
\end{figure*}

\subsection{Identification of the Lagrangian fronts favourable for saury fishing
in the season with the Second Oyashio Intrusion}

The oceanographic situation in the region cardinally differs in the years
with the Second Oyashio Intrusion
when a large anticyclonic eddy, formed as a warm-core Kuroshio ring,
approaches to the Hokkaido eastern coast and forces the
Oyashio Current to shift to the east rounding the eddy.
The SST image in Fig.~\ref{fig2}a averaged for 16--18 October 2004
demonstrates that the fishery
grounds have been concentrated mainly at the convergence front between the
Oyashio and Soya waters and the periphery of the anticyclonic
Kuroshio ring with the center
around $x_0=147^\circ 30'$E, $y_0=42^{\circ}30'$N.

The vorticity map in Fig.~\ref{fig2}b demonstrates that the maximal saury
catch spots are concentrated around the northeastern edge of the
warm anticyclonic Kuroshio ring, the big red spot with the center
around $x_0=147^\circ 30'$E, $y_0=42^{\circ}30'$N. The zonal, meridional,
and absolute drift maps on the same date
in Figs.~\ref{fig2}c, d, and e with the circles of saury catch locations
overlaid show clearly that the fishery grounds
with maximal catches are located along the LFs where productive cold waters
of the Oyashio Current, warmer waters of the southern branch of the Soya
Current and warm and salty waters of the Kuroshio ring converge.
The FTLE map in Fig.~\ref{fig2}f on 17 October 2004 demonstrates that
the fishery grounds are located along the three main unstable manifolds
in the region which demarcate approximately the corresponding LFs. Animation of the daily
Lagrangian maps for August~-- December
2004 with the fishery grounds overlaid is available at
http://dynalab.poi.dvo.ru/data/GRL12/2004.

\section{Frequency distribution of the distances between fishing boats with catches and Lagrangian fronts}

We demonstrated above a relationship between LFs and locations of fishing boats with saury catches
in different oceanographic situations.
In order to determine quantitatively whether saury is actively associating with the LFs in the region studied or not,
we compute the frequency distribution of the distances, $r$, between locations of fishing boats with saury catches
and the strong LFs for available fishery seasons. The maximal gradient of the absolute displacement,
$\nabla D=\sqrt{(\partial D/\partial x)^2 + (\partial D/\partial y)^2}$, is supposed to
be an identificator of the presence of a LF. Such gradients delineate boundaries between waters that passed
distances that may differ in two orders of magnitude (see Figs.~\ref{fig1}e and \ref{fig2}e). It has been shown in Sec.~4 that the contrast
boundaries
are the strongest LFs separating productive cold waters of the Oyashio Current, warmer waters of the southern branch
of the Soya Current, and waters of warm-core Kuroshio rings. In order to get rid of ephemeral LFs, we choose
a threshold, $\nabla D_{\rm th}=60$, and only the LFs with $\nabla D_{\rm th} \ge 60$ are supposed to be ``strong''.
We have found that such gradient values correspond to permanent LFs in the Kuroshio--Oyashio frontal area.
Then we compute on each day the distance $r$ between the location of each boat with 
saury catch and the nearest geographical point where $\nabla D_{\rm th} \ge 60$. 
The corresponding probability distribution function (PDF) for each season is compared with
the random PDF which is computed by the same way but with 10000 points randomly distributed over the same region.
The fishery strategy depends on the oceanographic situation
in the region, captain's experience, other subjective factors and varies during the fishery period. As a rule,
the Russian captains begin to catch saury in July -- August in the northern part of the region studied where saury schools
start their migration to the south. It is profitable to begin the fishery 
nearby the ports on the southern Kuril Islands.
The fishing grounds there are connected rather with the strong upwelling due to high tides than with LFs.
Then, in September, the fishing boats moved to the south following the saury schools. By safe and another reasons,
the boats prefer to stick together. After reporting by one of them a good catch, the other ones may move to that place
if they were hereabout.
The combination of different factors in the fishery strategy, including the subjective ones, makes it difficult
to find statistically significant correlations between locations of fishing boats
with saury catches and any oceanic features. After all, even if we were able to find precisely potential fishing grounds
with favourable hydrological conditions, it would not necessarily implies good real catching if saury or fishing boats
do not reach the place.

\begin{figure*}[htb]
\begin{center}
\includegraphics[width=0.8\textwidth,clip]{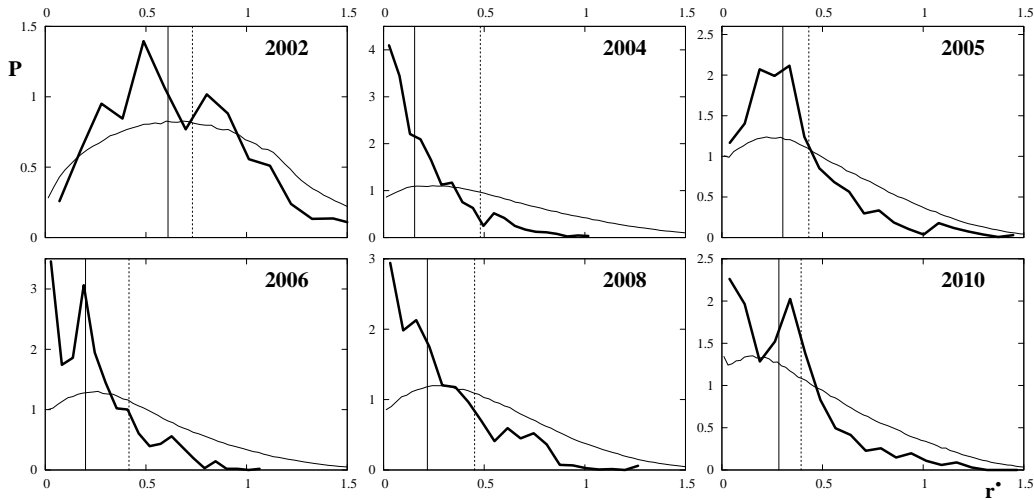}
\end{center}
\caption{PDFs for six fishery seasons with real locations of the fishing boats
with saury catches (solid curves) and the points randomly distributed over the same region (dashed curves).}
\label{fig3}
\end{figure*}

\begin{figure}[htb]
\begin{center}
\includegraphics[width=0.45\textwidth,clip]{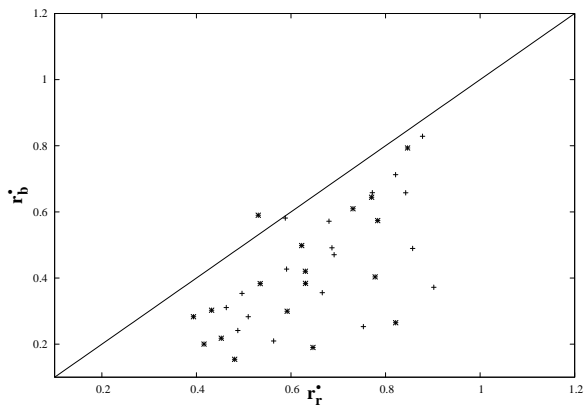}
\end{center}
\caption{Median (stars) and mean (crosses) values for locations of real boats, $r_b$, vs randomly distributed ones, $r_r$.}
\label{fig4}
\end{figure}

The results of our statistical analysis are shown in Fig.~\ref{fig3} for all available fishery seasons.
The number of events, i.~e., the number of locations of the boats with saury catches, varies from season to season
and in average was about 1000 per season. As expected, the random PDFs (thin curves) are rather
smooth curves with long tails. The real PDFs (bold curves) have a tooth-like structure that can be explained partly by
congregation of boats near strong LFs with large $D$ gradients and partly by the fishery strategy to stick together.
The vertical solid and dashed lines represent the medians for real and random PDFs, respectively.
By definition, the median of a finite list of numbers can be found by arranging all the observations from lowest value to
highest one and picking the middle one. It is such a location value that there exists in the list the same number
of locations smaller and larger than the median value.
The median is a more robust statistical indicator than the mean value and can be used as a measure of location
when a distribution is skewed having, for example, a heavy tail.
In all the seasons the medians were closer to the LFs for the real PDFs than for
the corresponding random ones. Moreover, the random PDFs have more
longer tails than the corresponding real ones proving that the fishing boats 
really tend to be closer to the LFs
then be randomly distributed over the region. The plot in Fig.~\ref{fig4} demonstrates the relations between medians (stars)
for the real and random PDFs and between mean values (crosses) for the real and random PDFs.
In order to take into account the effect of a choice of the threshold value for the displacement gradients,
$\nabla D_{\rm th}$, we compute the median and mean values for all the seasons at $\nabla D_{\rm th}=$60, 100, and 130.
The points below the slope line in Fig.~\ref{fig4} give evidence that the corresponding median or mean value is closer to LFs
for the real fishing boats, $r_b$, than for the randomly distributed ones, $r_r$. It is evident that both the medians and
mean values are closer to LFs in the first case.

The oceanographic situation in 2002 was not typical being the single fishery season, among the studied ones,
with the First Oyashio Intrusion. Moreover, the oceanographic situation in the region have changed significantly
during that season. Animation of daily Lagrangian maps in 2002
with the fishery grounds overlaid (http://dynalab.poi.dvo.ru/data/GRL12/2002) demonstrates clearly
that in addition to the intrusion of Soya Current waters along the north-eastern coast of Hokkaido island
there appeared in October -- November the intrusion of cold Oyashio waters from the north into the fishery region
which cardinally changed the oceanographic situation. The fishery grounds moved to the east and south in that time.
The most stable oceanographic situation among the years with available fishery data was in the year 2004
with the Second Oyashio Intrusion, when the quasistationary
warm Kuroshio anticyclonic ring was situated in the region for the whole fishery period. The strong LFs have been found to be
the permanent ones in that year. The real PDF in Fig.~\ref{fig3} in  2004 exceeds significantly the random one and decays rapidly.
The tail of the random PDF extends significantly over the distance $r$ as compared with the real one.
All these facts prove that saury fishing grounds really were located mainly along the strong LFs in that year.
In the other years with available fishery data, the oceanographic situations resemble
the 2004 case with the Second Oyashio Intrusion.
Probability distribution functions for those fishery seasons in Fig.~\ref{fig3} provide the statistical evidence that saury fishing locations
are not randomly distributed over the region but are concentrated near 
the strongest LFs around the Kuroshio ring (see Fig.~\ref{fig2}).
Based on statistical results, we may conclude that the more stable the oceanographic situation in the fishery region is,
the closer fishing boats to LFs tend to be.

Oceanic fronts are areas with strong horizontal and vertical mixing.
Highly turbid waters, however, are unsuitable for saury because it is a visual predator hunting in comparatively
clear waters outside the exact locations of fronts. They avoid highly turbid waters and waters with
large phytoplankton concentration, more than 5 g/m${}^3$, which are turbid due to organic matter.
On the other hand, extremely oligotrophic waters contain little food.
Food abundance and water clarity are known to be two factors affecting the rate of food encounter
\citep{Fukushima79,Saitoh86,Sugimoto92,Yasuda96}.

As to physical and biological reasons that may cause saury aggregation near LFs, we suggest the following ones.
Lagrangian Fronts in the Kuroshio--Oyashio frontal area demarcate the convergence of water masses with different
productivity. They are zones with increased lateral and vertical mixing and often with increased primary and secondary
production.  Mixture of nutrient rich Oyashio waters with more oligotrophic Kuroshio waters
can locally simulate phytoplankton photosynthesis and thus sustains higher phyto- and zooplankton
concentrations with a net effect of aggregation of saury to forage on the lower trophic level organisms.
Stretching of material lines in the vicinity of hyperbolic objects in the ocean, a hallmark of chaotic advection
(for a review see \citep{Wiggins05,Koshel06}), is one of the possible mechanisms providing effective intrusions of nutrient rich Oyashio cold
waters into more oligotrophic Kuroshio warm waters. Those filament-like intrusions may expand over hundreds of
kilometers and are easily captured by the Lagrangian diagnostics but may be not visible on SST or chlorophyll-a images.

\begin{figure}[htb]
\begin{center}
\includegraphics[width=0.45\textwidth,clip]{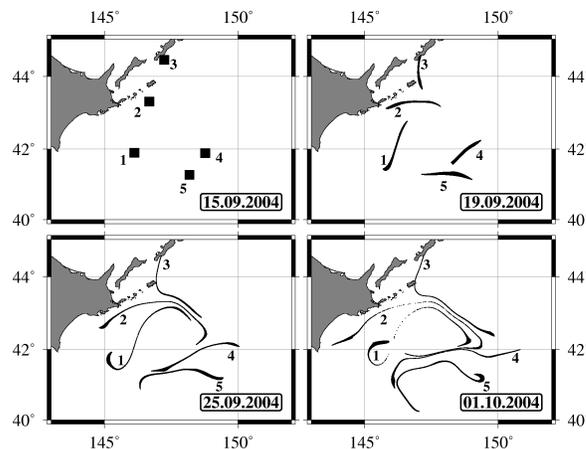}
\end{center}
\caption{Evolution of the synthetic food patches selected at 5 hyperbolic trajectories
in the 2004 fishery season with the Second Oyashio Intrusion.
For two weeks the patches delineate rapidly the corresponding unstable manifolds (the black ridges on the
FTLE map in Fig.~\ref{fig2}f) around the Kuroshio warm ring.}
\label{fig5}
\end{figure}

We illustrate that mechanism of transport of nutrients in
Fig.~\ref{fig5} where evolution of the patches with synthetic particles selected on September 15, 2004 at 5
hyperbolic trajectories in the region is shown. It is the fishery season with the Second Oyashio Intrusion and a
prominent quasistationary Kuroshio warm anticyclonic ring in the region (Fig.~\ref{fig2}). Let us suppose that
some of the patches are rich in food and trace their evolution. Comparing Fig.~\ref{fig5} with the backward-time 
FTLE map in Fig.~\ref{fig2}f, it becomes clear that all the patches in the course of time
delineate the corresponding ridges on the map which approximate the unstable manifolds of selected 5 hyperbolic
trajectories. The passive marine organisms in those fluid patches have been advected along
with them attracting saury for feeding. The perimeter of some patches increased more than in 100 times
for only two weeks increasing significantly the chance for saury to find food. Such a mechanism of export of 
nutrient rich waters into more poor ones is supposed to be typical 
because of a large number of hyperbolic trajectories in the frontal oceanic zones with increased mixing activity and 
rich in eddies.
Generally speaking, LFs are 3D features. Lagrangian maps allow to visualize their manifestations 
at the ocean surface. They may expand over hundreds of kilometers, but fish is expected to aggregate only near comparatively 
small LF segments. Saury, as a predator of zooplankton, prefer the places with an aggregation 
of forage zooplankton. Favourable fishing grounds may differ from the adjacent ones by a complex of 
conditions including hydrological situation in the upper ocean layer. 
It has been empirically found \citep{Filatov} that there should be a sharp 
seasonal termocline (with the vertical gradient more than 
0.19$^\circ$~C/m), coinciding in that region with  
a seasonal pycnocline. Moreover, the width of the upper mixed layer should be more 
than 6~m with the forage zooplankton concentration more than 
0.4 g/m${}^3$ but the phytoplankton concentration less than 5.6 g/m${}^3$. 
Such conditions are expected to be formed at LFs where water masses of 
different 
densities meet. The heavy cold water tends to flow under the light warm water 
resulting in 
formation of a sharp pycnocline which in turn force passive phytoplankton to 
concentrate all time near the surface 
but not to be distributed over the depth. Zooplankton is able to move 
vertically concentrating near the surface in the 
nighttime. The absolute values of SST in the locations with saury aggregation 
may vary from 5 to $20^{\circ}$~C. The impoprtant factors of favourable fishery 
conditions are thermostructural peculiarities and the adundance in forage zooplankton but not SST, salinity and other 
hydrographic factors.  

In the end of this section we would like to discuss the connection between potential fishing grounds,
LFs and thermal fronts visible
on SST images. SST fronts have long been the main indicators used to find places in the ocean with rich marine
resources. So, in order to find potential fishing grounds, it is instructive to use as the first guess the strong thermal
gradients visible on satellite SST images if they are available. Unfortunately, SST images are not available in cloudy
and rainy days which often occur in the fishery period in the region studied. In some fishery seasons
in the Kuroshio--Oyashio frontal area up to half of days
have been found to be cloudy or rainy. Typically, the saury fishing grounds with maximal catches
have been found not exactly at the SST fronts but in a comparatively large area around that front where a few LFs
may be detected. Computation of LFs is a simple way to visualize a fine structure of the frontal zone which is a problem
when using SST and/or chlorophyll-a images.
Any strong large-scale LFs can be accurately detected in a given altimetric geostrophic velocity field by computing synoptic
maps of the drift of synthetic tracers.
Moreover, by computing meridional and zonal drift maps, one gets an information on the origin and history
of convergent waters that may be useful to determine by which waters this or that LF has been formed.
Thus, the LFs, that can always be computed with AVISO altimetric geostrophic
velocity fields, may serve additional indicators of potential fishing grounds along with satellite SST images.

\section{Conclusions}

What is new we propose in this paper is a method to identify and analyze 
specific oceanic features in
altimetric (or given by another way) velocity fields that may serve as 
identificators of potential fishing grounds. 
We introduced the notion of a Lagrangian front (LF),
the region where surface waters with different origin and history converge, 
and showed how to find such fronts 
computing the zonal, meridional, and absolute drift maps for a large number of
synthetic particles in a given region.

Based on satellite-derived surface velocities, we have integrated advection equations for a large number of
synthetic tracers backward in time and computed the vorticity and FTLE maps, zonal,
meridional and absolute drift maps in the region to the east off the Hokkaido and southern Kuril Islands coasts,
one of the richest fishery place in the world. The data on fishing locations and daily catches of the Russian ships
were imposed on the SST, vorticity, FTLE and drift daily maps.
To determine quantitatively whether saury was actively associating with the LFs in the region studied or not,
we computed the frequency distribution of the distances between locations of fishing boats with saury catches
and the strong LFs for available fishery seasons. It has been shown statistically that the saury
fishing grounds with catches were not randomly distributed over the region but located mainly along those
LFs where productive cold waters of the Oyashio Current, warmer waters of the southern branch of
the Soya Current, and waters of warm-core Kuroshio rings converged.
We proposed a mechanism of effective export of nutrient rich waters into more poor ones based on stretching of material
lines in the vicinity of hyperbolic objects in the ocean. Those filament-like intrusions may expand over hundreds of
kilometers and are easily captured by the Lagrangian diagnostics but not visible on SST or chlorophyll-a images.
Thus, it has been shown that the strong LF locations may serve good indicators of
potential fishing grounds in rather different oceanographic conditions.

The method proposed seems to be quite general and may be applied to forecast
potential fishing grounds for the other pelagic fishes in different regions of the
World Ocean. On the other hand, our ability to recognize areas where pelagic fishes and
marine animals prefer to congregate may help to create protectable marine reservations there.
The Lagrangian tools can be useful with this aim.

\begin{acknowledgments}
This work was supported  by the Russian Foundation for Basic Research
(project nos. 11--05--98542, 12--05--00452, and 13--05--00099). The altimeter products
were distributed by AVISO with support from CNES. We would like to thank two anonymous reviewers  
for their very valuable suggestions that improved the quality of this paper.
\end{acknowledgments}

\section*{Appendix: Computation of the finite-time Lyapunov exponents via singular values of the evolution matrix}

The FTLE field characterizes quantitatively mixing along with directions of maximal stretching and contracting.
The LCS can be approximately found by computing local maxima (ridges) of the FTLE field. Ridges of the FTLE field reveal
stable manifolds when integrating advection equations (\ref{adveq}) forward in time
and unstable ones when integrating them backward in time. There are different methods
to compute FTLE (see, for a review \citep{Shadden05}). We use in this paper the method
introduced recently in Ref.~\citep{OM11} which is valid for $n$-dimensional vector fields
and enables to compute this quantity accurately even in very irregular velocity fields.
The Lyapunov exponents in this method are defined via singular values of the evolution matrix.

Let us represent the advection equations in a $n$-dimensional velocity field in the vector form
\begin{equation}
\begin{gathered}
\mathbf{\dot x}=\mathbf{f}(\mathbf{x},t),\quad
\mathbf{x}=(x_1,\dotsc,x_n),\\
\mathbf{f}(\mathbf{x},t)=(f_1(x_1,\dotsc,x_n,t),\dots,f_n(x_1,\dotsc,x_n,t)).
\end{gathered}
\label{nonlinsys}
\end{equation}
The Lyapunov exponent at an arbitrary point $\mathbf{x_0}$ is given by
\begin{equation}
\Lambda (\mathbf{x_0})=\lim_{t\to\infty}\lim_{\Vert\delta \mathbf{x}(0)\Vert\to 0}\frac{\ln(\Vert\delta \mathbf{x}(t)\Vert/\Vert\delta \mathbf{x}(0)\Vert)}{t},
\label{lyap_def}
\end{equation}
where $\delta \mathbf{x}(t)=\mathbf{x_1}(t)-\mathbf{x_0}(t)$ is an infinitesimally small distance,
$\mathbf{x_0}(t)$ and $\mathbf{x_1}(t)$~ are solutions of the set
(\ref{nonlinsys}), $\mathbf{x_0}(0)=\mathbf{x_0}$. The limit exists,
it is the same for almost all the choices of $\delta \mathbf{x}(0)$ and
has a clear geometrical sense: trajectories of two nearby particles
diverge in time exponentially (in average) with the rate given by
the Lyapunov exponent.

Due to smallness of $\delta \mathbf{x}$ one can linearize the set
(\ref{nonlinsys}) in a vicinity of some trajectory $\mathbf{x_0}(t)$
and obtain the system of time-dependent linear equations
\begin{equation}
\begin{pmatrix}
\delta\dot x_1\\
\hdotsfor{1}\\
\delta\dot x_n
\end{pmatrix}=J(t)
\begin{pmatrix}
\delta x_1\\
\hdotsfor{1}\\
\delta x_n
\end{pmatrix},
\label{linearization}
\end{equation}
where $J(t)$ is the Jacobian matrix of the system (\ref{nonlinsys}) along
the trajectory $\mathbf{x_0}(t)$.
Solution of the linear system (\ref{linearization}) can be found with the
help of the evolution matrix $G(t,t_0)$
\begin{equation}
\begin{pmatrix}
\delta x_1(t)\\
\hdotsfor{1}\\
\delta x_n(t)
\end{pmatrix}=G(t,t_0)
\begin{pmatrix}
\delta x_1(t_0)\\
\hdotsfor{1}\\
\delta x_n(t_0)
\end{pmatrix},
\label{evol_mat}
\end{equation}
that obeys the differential equation which can be obtained after substituting~(\ref{evol_mat})
into~(\ref{linearization})
\begin{equation}
\dot G=JG,
\label{evol_mat_diffur}
\end{equation}
with the initial condition $G(t_0,t_0)=I$, where $I$ is the unit matrix.
Any evolution matrix has the important property
\begin{equation}
G(t,t_0)=G(t,t_1)G(t_1,t_0).
\label{evol_mat_prop}
\end{equation}

One can write the singular-value decomposition  of the evolution matrix as follows:
\begin{equation}
G(t,t_0)=U(t,t_0)D(t,t_0)V^T(t,t_0),
\label{evol_mat_svd}
\end{equation}
where $U$, $V$ are orthogonal and $D=\operatorname{diag}(\sigma_1,\dots,
\sigma_n)$ is diagonal.
The quantities $\sigma_1,\dots,\sigma_n$ are called singular values of
the matrix $G$. The evolution matrix, $G$, transforms a sphere of the unit radius to the
ellipsoid with the semiaxes to be equal to $\sigma_1,\dots,\sigma_n$.
The Lyapunov exponents are defined via singular
values of the evolution matrix as follows:
\begin{equation}
\Lambda_i=\lim_{t\to\infty}\frac{\ln\sigma_i(t,t_0)}{t-t_0}.
\label{lyap_def_sigma}
\end{equation}
Quantities
\begin{equation}
\lambda_i(t,t_0)=\frac{\ln\sigma_i(t,t_0)}{t-t_0}
%\label{lyap_ftle}
\end{equation}
are called FTLE which is the ratio of the logarithm of the maximal possible
stretching in a given direction to a time interval $t-t_0$.

The formulae, derived up to now, are valid with any $n$-dimensional version of
the original ordinary nonlinear differential equations (\ref{nonlinsys}).
In the two-dimensional case of particle's advection on a surface
the singular-value decomposition of the $2\times 2$ evolution matrix is
as follows:
\begin{multline}
G=UDV^T \Rightarrow
\begin{pmatrix}
a&b\\c&d
\end{pmatrix}
=
\begin{pmatrix}
\cos\phi_2&-\sin\phi_2\\
\sin\phi_2&\cos\phi_2
\end{pmatrix}\times\\
\begin{pmatrix}
\sigma_1&0\\
0&\sigma_2
\end{pmatrix}
\begin{pmatrix}
\cos\phi_1&-\sin\phi_1\\
\sin\phi_1&\cos\phi_1
\end{pmatrix}.
\label{SVD2x2}
\end{multline}
Solution of these four algebraic equations are
\begin{equation}
\begin{gathered}
\sigma_1=\frac{\sqrt{(a+d)^2+(c-b)^2}+\sqrt{(a-d)^2+(b+c)^2}}{2},\\
\sigma_2=\frac{\sqrt{(a+d)^2+(c-b)^2}-\sqrt{(a-d)^2+(b+c)^2}}{2},\\
\phi_1=\frac{\operatorname{arctan2}{(c-b,\,a+d)}-\operatorname{arctan2}{(c+b,\,a-d)}}{2},\\
\phi_2=\frac{\operatorname{arctan2}{(c-b,\,a+d)}+\operatorname{arctan2}{(c+b,\,a-d})}{2},
\end{gathered}
\label{SVD2x2finsol}
\end{equation}
where function $\operatorname{arctan2}$ is defined as
\begin{equation}
\operatorname{arctan2}{(y,x)}=\left\{
\begin{aligned}
&\arctan{(y/x)}, &x\ge 0,\\
&\arctan{(y/x)}+\pi, &x<0.
\end{aligned}\right.
\label{arctg2}
\end{equation}

When integrating Eq.~(\ref{evol_mat_diffur}) numerically,
we divide a large time interval on subintervals with the duration
less or order of the Lyapunov time, $t_\lambda=1/\lambda$,
and represent the whole evolution matrix as a product of evolution matrices
computed on these subintervals using the property~(\ref{evol_mat_prop}).
We compute this product and the corresponding singular values using the
GNU Multiple Precision Arithmetic Library~(http://gmplib.org) in order
to preserve the absolute precision of our representation of the
evolution matrix.

\bibliography{paper}

\begin{thebibliography}{38}
\providecommand{\natexlab}[1]{#1}
\expandafter\ifx\csname urlstyle\endcsname\relax
  \providecommand{\doi}[1]{doi:\discretionary{}{}{}#1}\else
  \providecommand{\doi}{doi:\discretionary{}{}{}\begingroup
  \urlstyle{rm}\Url}\fi

\bibitem[{\textit{Abraham and Bowen}(2002)}]{Abraham02}
Abraham, E.~R., and M.~M. Bowen (2002), {Chaotic stirring by a mesoscale
  surface-ocean flow}, \textit{Chaos}, \textit{12}(2), 373--381,
  \doi{10.1063/1.1481615}.

\bibitem[{\textit{Bakun}(2006)}]{Bakun}
Bakun, A. (2006), {Fronts and eddies as key structures in the habitat of marine
  fish larvae: opportunity, adaptive response and competitive advantage},
  \textit{Scientia Marina}, \textit{70}(S2), 105--122,
  \doi{10.3989/scimar.2006.70s2105}.

\bibitem[{\textit{Beron-Vera et~al.}(2008)\textit{Beron-Vera, Olascoaga, and
  Goni}}]{Beron08}
Beron-Vera, F.~J., M.~J. Olascoaga, and G.~J. Goni (2008), {Oceanic mesoscale
  eddies as revealed by {L}agrangian coherent structures}, \textit{Geophys.
  Res. Lett.}, \textit{35}(12), L12603, \doi{10.1029/2008GL033957}.

\bibitem[{\textit{Boffetta et~al.}(2001)\textit{Boffetta, Lacorata, Redaelli,
  and Vulpiani}}]{Boffetta01}
Boffetta, G., G.~Lacorata, G.~Redaelli, and A.~Vulpiani (2001), {Detecting
  barriers to transport: a review of different techniques}, \textit{Physica D},
  \textit{159}(1--2), 58--70, \doi{10.1016/S0167-2789(01)00330-X}.

\bibitem[{\textit{d'Ovidio et~al.}(2004)\textit{d'Ovidio, Fern{\'{a}}ndez,
  Hern{\'{a}}ndez-Garc{\'{\i}}a, and L{\'{o}}pez}}]{Ovidio04}
d'Ovidio, F., V.~Fern{\'{a}}ndez, E.~Hern{\'{a}}ndez-Garc{\'{\i}}a, and
  C.~L{\'{o}}pez (2004), {Mixing structures in the {M}editerranean {S}ea from
  finite-size {L}yapunov exponents}, \textit{Geophys. Res. Lett.},
  \textit{31}(17), L17203, \doi{10.1029/2004GL020328}.

\bibitem[{\textit{d'Ovidio et~al.}(2009)\textit{d'Ovidio, Isern-Fontanet,
  L{\'{o}}pez, Hern{\'{a}}ndez-Garc{\'{\i}}a, and
  Garc{\'{\i}}a-Ladona}}]{Ovidio09}
d'Ovidio, F., J.~Isern-Fontanet, C.~L{\'{o}}pez,
  E.~Hern{\'{a}}ndez-Garc{\'{\i}}a, and E.~Garc{\'{\i}}a-Ladona (2009),
  {Comparison between {E}ulerian diagnostics and finite-size {L}yapunov
  exponents computed from altimetry in the {A}lgerian basin}, \textit{Deep Sea
  Res. Part I}, \textit{56}(1), 15--31, \doi{10.1016/j.dsr.2008.07.014}.

\bibitem[{\textit{Filatov}(2005)}]{Filatov}
Filatov, V.~N. (2005), Pacific saury migrations in the areas of the {K}uril
  {I}slands and the {S}ea of {O}khotsk, in \textit{Proc. 20th Int. Symp. on
  Okhotsk Sea and Sea Ice}, pp. 257--260.

\bibitem[{\textit{Fukushima}(1979)}]{Fukushima79}
Fukushima, S. (1979), {Synoptic analysis of migration and fishing conditions of
  saury in the northwest {P}acific {O}cean}, \textit{Bull. Tohoku Reg. Fish.
  Res. Lab.}, \textit{4}, 1--70, in Japanese, English abstract.

\bibitem[{\textit{Haller}(2000)}]{H00}
Haller, G. (2000), {Finding finite-time invariant manifolds in two-dimensional
  velocity fields}, \textit{Chaos}, \textit{10}(1), 99--108,
  \doi{10.1063/1.166479}.

\bibitem[{\textit{Haller}(2002)}]{H02}
Haller, G. (2002), {Lagrangian coherent structures from approximate velocity
  data}, \textit{Phys. Fluids}, \textit{14}(6), 1851, \doi{10.1063/1.1477449}.

\bibitem[{\textit{Haller}(2011)}]{H11}
Haller, G. (2011), {A variational theory of hyperbolic Lagrangian Coherent
  Structures}, \textit{Physica D}, \textit{240}(7), 574--598,
  \doi{10.1016/j.physd.2010.11.010}.

\bibitem[{\textit{Haller and Beron-Vera}(2012)}]{H12}
Haller, G., and F.~J. Beron-Vera (2012), {Geodesic theory of transport barriers
  in two-dimensional flows}, \textit{Physica D}, \textit{241}(20), 1680--1702,
  \doi{10.1016/j.physd.2012.06.012}.

\bibitem[{\textit{Haller and Poje}(1998)}]{Haller}
Haller, G., and A.~C. Poje (1998), {Finite time transport in aperiodic flows},
  \textit{Physica D}, \textit{119}(3--4), 352--380,
  \doi{10.1016/S0167-2789(98)00091-8}.

\bibitem[{\textit{Haller and Yuan}(2000)}]{Haller00}
Haller, G., and G.~Yuan (2000), {Lagrangian coherent structures and mixing in
  two-dimensional turbulence}, \textit{Physica D}, \textit{147}(3--4),
  352--370, \doi{10.1016/S0167-2789(00)00142-1}.

\bibitem[{\textit{Harrison and Glatzmaier}(2012)}]{Harrison10}
Harrison, C.~S., and G.~A. Glatzmaier (2012), {Lagrangian coherent structures
  in the California Current System -- sensitivities and limitations},
  \textit{Geophys. Astrophys. Fluid Dyn.}, \textit{106}(1), 22--44,
  \doi{10.1080/03091929.2010.532793}.

\bibitem[{\textit{Hern{\'{a}}ndez-Carrasco
  et~al.}(2011)\textit{Hern{\'{a}}ndez-Carrasco, L{\'{o}}pez,
  Hern{\'{a}}ndez-Garc{\'{\i}}a, and Turiel}}]{Hernandez11}
Hern{\'{a}}ndez-Carrasco, I., C.~L{\'{o}}pez, E.~Hern{\'{a}}ndez-Garc{\'{\i}}a,
  and A.~Turiel (2011), {How reliable are finite-size Lyapunov exponents for
  the assessment of ocean dynamics?}, \textit{Ocean Modelling},
  \textit{36}(3-4), 208--218, \doi{10.1016/j.ocemod.2010.12.006}.

\bibitem[{\textit{Huhn et~al.}(2012)\textit{Huhn, von Kameke,
  P{\'{e}}rez-Mu{\~{n}}uzuri, Olascoaga, and Beron-Vera}}]{Huhn12}
Huhn, F., A.~von Kameke, V.~P{\'{e}}rez-Mu{\~{n}}uzuri, M.~J. Olascoaga, and
  F.~J. Beron-Vera (2012), {The impact of advective transport by the {S}outh
  {I}ndian {O}cean {C}ountercurrent on the {M}adagascar plankton bloom},
  \textit{Geophys. Res. Lett.}, \textit{39}(6), L06602,
  \doi{10.1029/2012GL051246}.

\bibitem[{\textit{Keating et~al.}(2011)\textit{Keating, Smith, and
  Kramer}}]{Keating11}
Keating, S.~R., K.~S. Smith, and P.~R. Kramer (2011), {Diagnosing Lateral
  Mixing in the Upper Ocean with Virtual Tracers: Spatial and Temporal
  Resolution Dependence}, \textit{J. Phys. Oceanogr.}, \textit{41}(8),
  1512--1534, \doi{10.1175/2011JPO4580.1}.

\bibitem[{\textit{{Kirwan Jr.}}(2006)}]{Kirwan06}
{Kirwan Jr.}, A. (2006), {Dynamics of ``critical'' trajectories},
  \textit{Progress In Oceanography}, \textit{70}(2--4), 448--465,
  \doi{10.1016/j.pocean.2005.07.002}.

\bibitem[{\textit{Koshel' and Prants}(2006)}]{Koshel06}
Koshel', K.~V., and S.~V. Prants (2006), {Chaotic advection in the ocean},
  \textit{Phys.--Usp.}, \textit{49}(11), 1151--1178,
  \doi{10.1070/PU2006v049n11ABEH006066}.

\bibitem[{\textit{Lehahn et~al.}(2007)\textit{Lehahn, d'Ovidio, L{\'{e}}vy, and
  Heifetz}}]{Lehahn07}
Lehahn, Y., F.~d'Ovidio, M.~L{\'{e}}vy, and E.~Heifetz (2007), {Stirring of the
  northeast {A}tlantic spring bloom: {A} {L}agrangian analysis based on
  multisatellite data}, \textit{J. Geophys. Res.}, \textit{112}(C8), C08005,
  \doi{10.1029/2006JC003927}.

\bibitem[{\textit{Lukovich and Shepherd}(2005)}]{Lukovich}
Lukovich, J.~V., and T.~G. Shepherd (2005), {Stirring and Mixing in
  Two-Dimensional Divergent Flow}, \textit{J. Atmos. Sci.}, \textit{62}(11),
  3933--3954, \doi{10.1175/JAS3580.1}.

\bibitem[{\textit{Mancho et~al.}(2004)\textit{Mancho, Small, and
  Wiggins}}]{Mancho04}
Mancho, A.~M., D.~Small, and S.~Wiggins (2004), {Computation of hyperbolic
  trajectories and their stable and unstable manifolds for oceanographic flows
  represented as data sets}, \textit{Nonlin. Proc. Geophys.}, \textit{11}(1),
  17--33, \doi{10.5194/npg-11-17-2004}.

\bibitem[{\textit{Mendoza and Mancho}(2010)}]{Mancho2009}
Mendoza, C., and A.~M. Mancho (2010), {Hidden Geometry of Ocean Flows},
  \textit{Phys. Rev. Lett.}, \textit{105}(3), 038,501,
  \doi{10.1103/PhysRevLett.105.038501}.

\bibitem[{\textit{Olascoaga et~al.}(2006)\textit{Olascoaga, Rypina, Brown,
  Beron-Vera, Ko{\c{c}}ak, Brand, Halliwell, and Shay}}]{Olascoaga06}
Olascoaga, M.~J., I.~I. Rypina, M.~G. Brown, F.~J. Beron-Vera, H.~Ko{\c{c}}ak,
  L.~E. Brand, G.~R. Halliwell, and L.~K. Shay (2006), {Persistent transport
  barrier on the {W}est {F}lorida {S}helf}, \textit{Geophys. Res. Lett.},
  \textit{33}(22), L22603, \doi{10.1029/2006GL027800}.

\bibitem[{\textit{Olson et~al.}(1994)\textit{Olson, Hitchcock, Mariano,
  Ashjian, Peng, Nero, and Podesta}}]{Olson}
Olson, D., G.~Hitchcock, A.~Mariano, C.~Ashjian, G.~Peng, R.~Nero, and
  G.~Podesta (1994), {Life on the Edge: Marine Life and Fronts},
  \textit{Oceanography}, \textit{7}(2), 52--60, \doi{10.5670/oceanog.1994.03}.

\bibitem[{\textit{Owen}(1981)}]{Owen}
Owen, R. (1981), {Fronts and Eddies in the Sea: Mechanisms, Interactions and
  Biological Effects}, in \textit{{Analysis of marine ecosystems}}, edited by
  A.~R. Longhurst, pp. 197--233, Academic Press Inc., London.

\bibitem[{\textit{Prants et~al.}(2011{\natexlab{a}})\textit{Prants, Budyansky,
  Ponomarev, and Uleysky}}]{OM11}
Prants, S., M.~Budyansky, V.~Ponomarev, and M.~Uleysky (2011{\natexlab{a}}),
  {Lagrangian study of transport and mixing in a mesoscale eddy street},
  \textit{Ocean Modelling}, \textit{38}(1--2), 114--125,
  \doi{10.1016/j.ocemod.2011.02.008}.

\bibitem[{\textit{Prants et~al.}(2011{\natexlab{b}})\textit{Prants, Uleysky,
  and Budyansky}}]{DAN11}
Prants, S.~V., M.~Y. Uleysky, and M.~V. Budyansky (2011{\natexlab{b}}),
  {Numerical simulation of propagation of radioactive pollution in the ocean
  from the {F}ukushima {D}ai-ichi nuclear power plant}, \textit{Doklady Earth
  Sciences}, \textit{439}(2), 1179--1182, \doi{10.1134/S1028334X11080277}.

\bibitem[{\textit{Prants et~al.}(2013)\textit{Prants, Ponomarev, Budyansky,
  Uleysky, and Fayman}}]{FAO13}
Prants, S.~V., V.~I. Ponomarev, M.~V. Budyansky, M.~Y. Uleysky, and P.~A.
  Fayman (2013), {Lagrangian analysis of mixing and transport of water masses
  in the marine bays}, \textit{Izvestiya, Atmospheric and Oceanic Physics},
  \textit{49}(1), 82--96, \doi{10.1134/S0001433813010088}.

\bibitem[{\textit{Rypina et~al.}(2011)\textit{Rypina, Scott, Pratt, and
  Brown}}]{Rypina11}
Rypina, I.~I., S.~E. Scott, L.~J. Pratt, and M.~G. Brown (2011), {Investigating
  the connection between complexity of isolated trajectories and Lagrangian
  coherent structures}, \textit{Nonlin. Proc. Geophys.}, \textit{18}(6),
  977--987, \doi{10.5194/npg-18-977-2011}.

\bibitem[{\textit{Saitoh et~al.}(1986)\textit{Saitoh, Kosaka, and
  Iisaka}}]{Saitoh86}
Saitoh, S., S.~Kosaka, and J.~Iisaka (1986), {Satellite infrared observations
  of {K}uroshio warm-core rings and their application to study of {P}acific
  saury migration}, \textit{Deep Sea Res. Part A.}, \textit{33}(11--12),
  1601--1615, \doi{10.1016/0198-0149(86)90069-5}.

\bibitem[{\textit{Shadden et~al.}(2005)\textit{Shadden, Lekien, and
  Marsden}}]{Shadden05}
Shadden, S.~C., F.~Lekien, and J.~E. Marsden (2005), {Definition and properties
  of {L}agrangian coherent structures from finite-time {L}yapunov exponents in
  two-dimensional aperiodic flows}, \textit{Physica D}, \textit{212}(3--4),
  271--304, \doi{10.1016/j.physd.2005.10.007}.

\bibitem[{\textit{Sugimoto and Tameishi}(1992)}]{Sugimoto92}
Sugimoto, T., and H.~Tameishi (1992), {Warm-core rings, streamers and their
  role on the fishing ground formation around {J}apan}, \textit{Deep Sea Res.
  Part A.}, \textit{39, Supplement 1}, S183--S201,
  \doi{10.1016/S0198-0149(11)80011-7}.

\bibitem[{\textit{{Tew Kai} et~al.}(2009)\textit{{Tew Kai}, Rossi, Sudre,
  Weimerskirch, Lopez, Hernandez-Garcia, Marsac, and Gar{\c{c}}on}}]{Kai09}
{Tew Kai}, E., V.~Rossi, J.~Sudre, H.~Weimerskirch, C.~Lopez,
  E.~Hernandez-Garcia, F.~Marsac, and V.~Gar{\c{c}}on (2009), {Top marine
  predators track {L}agrangian coherent structures}, \textit{Proc. of the
  National Academy of Sciences of the USA}, \textit{106}(20), 8245--8250,
  \doi{10.1073/pnas.0811034106}.

\bibitem[{\textit{Uda}(1938)}]{Uda38}
Uda, M. (1938), {Researches on ``{S}iome'' or current rip in the seas and
  oceans}, \textit{Geophys. Mag.}, \textit{11}, 307--372.

\bibitem[{\textit{Wiggins}(2005)}]{Wiggins05}
Wiggins, S. (2005), {The dynamical systems approach to {L}agrangian transport
  in oceanic flows}, \textit{Annu. Rev. Fluid Mech.}, \textit{37}(1), 295--328,
  \doi{10.1146/annurev.fluid.37.061903.175815}.

\bibitem[{\textit{Yasuda and Kitagawa}(1996)}]{Yasuda96}
Yasuda, I., and D.~Kitagawa (1996), {Locations of early fishing grounds of
  saury in the northwestern {P}acific}, \textit{Fisheries Oceanography},
  \textit{5}(1), 63--69, \doi{10.1111/j.1365-2419.1996.tb00018.x}.

\end{thebibliography}
\bibliographystyle{agu08}

\end{article}

\end{document}